\documentclass[10pt]{iopart}
\usepackage{graphicx}

\begin{document}

\title{Analysis of finite-size effect of infinite-range Ising model under Glauber dynamics}

\author{Hisato Komatsu } 
\address{Research Center for Advanced Measurement and Characterization, National Institute for Materials Science, Tsukuba, Ibaraki 305-0047, Japan}
\ead{KOMATSU.Hisato@nims.go.jp}

\begin{abstract}

We consider an infinite-range Ising model under the Glauber dynamics and determine the finite-size effect on the distribution of two spin variables as a perturbation of $O \left( 1/N \right)$. Based on several considerations, ordinary differential equations are derived for describing the time development of both a two-body correlation and the autocorrelation function of magnetization. The results of the calculation fit the simulation results, unless the perturbation theory breaks down because of critical phenomena or magnetization reversal.

\end{abstract}

\maketitle

\section{Introduction \label{Introduction} }

Finite-size effect has an important role in various problems of statistical physics such as the scaling analysis of critical phenomena, not only occurring in equilibrium states but also in nonequilibrium ones. However, its consideration in nonequilibrium systems is generally more difficult than in equilibrium ones. Hence, an infinite-range Ising model under the Glauber dynamics, one of the simplest models of such systems, has been studied as an example\cite{PHB89,AFC10,MMR10,GMM11}. Most of these studies considered the probability density function of magnetization, $P_{\mathrm{whole} } (m' )$, and used the Fokker--Planck equation describing the time development of this function. To derive this equation, they first calculated the Kramers--Moyal expansion corresponding to the master equation and ignored the terms containing higher-order derivatives, considering them as higher-order infinitesimals. Similar manipulation of the master equation is studied also in other contexts, such as chemical reactions\cite{vK76,HGTT84,LT10,PT18}.
However, to express the Kramers--Moyal expansion, the number of microscopic states under fixed-order parameters needs to be determined. Hence, it is difficult to generalize the above method to other types of infinite-range models, and other approaches are required for these models.

In this study, we consider the probability distribution of two spin variables $P_2 ( \sigma _i , \sigma _j ; t )$ and calculate two-body correlation $\left< \delta \sigma _i \delta \sigma _j \right>$ and autocorrelation of magnetization $C(t_0 , t) $. The finite-size effect on them is treated as a $ O \left( 1/N \right)$ perturbation. Note that the perturbation terms of these properties play significant roles because the spins of the infinite-range model are effectively independent of each other in the thermodynamic limit. After several considerations, the time development of these properties is expressed by ordinary differential equations of several parameters. In our method, we assume that the probability distribution of magnetization $M$ is approximated as a Gaussian one; therefore, the third-order cumulant of $M$ is approximately zero. This assumption is necessary for avoiding the problem called BBGKY hierarchy\cite{CDFR14}. To confirm the validity of the perturbation theory and this assumption, the results of the derived differential equations are compared with those of numerical simulations. Furthermore, we prove that the differential equations describing magnetization and the two-body correlation are also derived from the Fokker--Planck equation under the assumption of a Gaussian distribution for magnetization (see the appendix). Note that although our method and the Fokker--Planck equation give equivalent results, the former has an advantage in that it can be applied to other infinite-range models more easily than the latter.

 The remainder of this paper is organized as follows. First, we explain the model in section \ref{Model}, review its behaviour in the thermodynamic limit in section \ref{Ninf}, and present the calculation of the time development of $ \left< \delta \sigma _i \delta \sigma _j \right>$ and $C(t_0 , t) $ in section \ref{2body} and \ref{auto_corr}, respectively. Finally, the study is summarized in section \ref{Summary}. In \ref{App1}, we prove that our method and the Fokker--Planck equation yield the same conclusion, at least in the case of $ \left< \delta \sigma _i \delta \sigma _j \right>$.

\section{Model \label{Model} }

We consider the following Ising model with an infinite-range interaction:
\begin{equation}
{\cal H}  =  - \frac{J}{N} \sum _{ i , j } \sigma _{i} \sigma _{j} - h \sum _i \sigma _i = - \frac{J}{N} \left( \sum _{i} \sigma _{i} \right) ^2 - h \sum _i \sigma _i , \label{Hamiltonian}
\end{equation}
\begin{equation}
\mathrm{where} \ \ \ \sigma _i =  \pm 1 , \label{Ising_spin}
\end{equation}
and consider the dynamics using the Markov chain Monte Carlo (MCMC) method. In this study, for simplicity, we consider a case that there is no magnetic field, i.e., $h=0$. Here, the updating of each step is the flipping of one randomly chosen spin $\sigma _i$, i.e.,
\begin{equation}
\left\{ \sigma \right\} = (\sigma _1 , \sigma _2 , ... , \sigma _N ) \mapsto \left\{ F_i \sigma \right\} \equiv (\sigma _1 , \sigma _2 , ... , - \sigma _i , ... , \sigma _N ) ,
\end{equation}
and we define the unit of time $t$ as 1 Monte Carlo step (MCS). The time development of the probability distribution of the spin configuration, $P_N ( \left\{ \sigma \right\} ; t )$, is expressed as follows:
\begin{eqnarray}
 P_N \left( \left\{ \sigma \right\} ; t+ \frac{1}{N} \right) & = & P_N ( \left\{ \sigma \right\} ; t ) \nonumber \\
& & + \frac{1}{N} \sum _i \bigl\{ P_N ( \left\{ F_i \sigma \right\} ; t ) W \left( \left\{ F_i \sigma \right\} \rightarrow \left\{ \sigma \right\} \right) \bigr. \nonumber \\
& & \bigl. - P_N ( \left\{ \sigma \right\} ; t ) W( \left\{ \sigma \right\} \rightarrow \left\{ F_i \sigma \right\} ) \bigr\} ,
 \label{MF_dynam1}
\end{eqnarray}
where $W( \left\{ \sigma \right\} \rightarrow \left\{ \sigma ' \right\} )$ is the acceptance ratio of the updating, $ \left\{ \sigma \right\} \rightarrow \left\{ \sigma ' \right\} $. Note that $1/N$ MCS implies one step of updating. In the typical MCMC method, $W$ is a function of the product of the energy change during the updating and the inverse temperature.
\begin{equation}
\beta \delta E ( \left\{ \sigma \right\} \rightarrow \left\{ F_i \sigma \right\} ) = -\frac{2\beta J}{N} \cdot ( - \sigma _i - \sigma _i ) \sum _{ j \neq i } \sigma _j = 4 \beta J \sigma _i \frac{ \sum _{ j \neq i } \sigma _j }{N} ; 
\end{equation}
therefore, we can express that 
\begin{equation}
 W( \left\{ \sigma \right\} \rightarrow \left\{ F_i \sigma \right\} ) = w \left(  4 \beta J \sigma _i \frac{ \sum _{ j \neq i } \sigma _j }{N} \right) .
 \label{transition0}
\end{equation}
There are several ways for defining the form of $w$. In this study, we adopt the Glauber dynamics as follows:
\begin{equation}
w( x ) = \frac{1}{1 + e^x } = \frac{1 - \tanh \left( \frac{x}{2} \right) }{2} .
\label{transition1}
\end{equation}
When the system size, $N$, is large, the average of the spins, $ \left( \sum _{ j \neq i } \sigma _j \right)/N$, can be approximated as their ensemble average $m = \left< \sigma _i \right>$ with an $O(1/\sqrt{N})$ fluctuation, i.e.,  
\begin{equation}
\frac{ \sum _{ j \neq i } \sigma _j }{N} = m + O \left( \frac{1}{\sqrt{N} }  \right)
\label{m_fluctuation}
\end{equation}
Substituting the above relation into (\ref{transition0}), we obtain the Taylor expansion of $W$ as
\begin{eqnarray}
 W( \left\{ \sigma \right\} \rightarrow \left\{ F_i \sigma \right\} ) & = & w \left(  4 \beta J m \sigma _i  \right) + 4 \beta J \sigma _i \left( \frac{\sum _{j \neq i} \sigma _j }{N} -m \right) w' \left( 4 \beta J m \sigma _i  \right) \nonumber \\
& & + 8 \left( \beta J \right) ^2 \left( \frac{\sum _{j \neq i} \sigma _j }{N} -m \right) ^2 w'' \left( 4 \beta J m \sigma _i  \right) + o \left( \frac{1}{N} \right) . \label{transition2}
\end{eqnarray}

\section{Calculations }

In this section, we present the calculation of the dynamics under the transition rate expressed in (\ref{transition2}). In section \ref{Ninf}, first, we survey the dynamics in the case of an extremely large $N$ as the zeroth approximation of the finite-$N$ system. Subsequently, in sections \ref{2body} and \ref{auto_corr}, we consider an $O \left( 1/N \right)$ modification of the two-body correlation, $\left< \delta \sigma _i \delta \sigma _j \right> $, and the autocorrelation of magnetization, $C(t_0 , t) $, respectively. 

\subsection{Dynamics when $N$ is large \label{Ninf} } 

In the case $N$ is extremely large, the second and third terms of (\ref{transition2}) converge to zero; therefore, (\ref{MF_dynam1}) can be expressed as
\begin{eqnarray}
 P_N \left( \left\{ \sigma \right\} ; t+ \frac{1}{N} \right) & = & P_N ( \left\{ \sigma \right\} ; t ) \nonumber \\
& & + \frac{1}{N} \sum _i \bigl\{ P_N ( \left\{ F_i \sigma \right\} ; t ) w \left( - 4 \beta J m \sigma _i  \right) - P_N ( \left\{ \sigma \right\} ; t ) w \left(  4 \beta J m \sigma _i  \right)  \bigr\} .
 \label{MF_dynam2}
\end{eqnarray}
First, we define $P_s$ as
\begin{equation}
 P_s ( \sigma _{i_1} , \sigma _{i_2} , ... , \sigma _{i_s} ; t ) \equiv \mathrm{Tr} _{ \left\{ \sigma _n \right\} _{n \neq i_1 , i_2 , ... , i_s } } P_N ( \left\{ \sigma \right\} ; t ) ,
 \label{Ps0}
\end{equation}
and derive its time development under (\ref{MF_dynam2}). Taking the summation over all spins except $\sigma _{i_1} , \sigma _{i_2} , ... , \sigma _{i_s}$, (\ref{MF_dynam2}) can be transformed into an equation describing the time development of $P_s$.
\begin{eqnarray}
 P_s \left( \sigma _{i_1} , \sigma _{i_2} , ... , \sigma _{i_s} ; t + \frac{1}{N} \right) & = & P_s ( \sigma _{i_1} , \sigma _{i_2} , ... , \sigma _{i_s} ; t ) \nonumber \\
& & + \frac{1}{N} \sum _{n=1} ^s \bigl\{ P_s ( \sigma _{i_1} , \sigma _{i_2} , ... , -\sigma _{i_n} , ... , \sigma _{i_s} ; t ) w \left( - 4 \beta J m \sigma _{i_n}  \right) \biggr. \nonumber \\
& & \biggl.  - P_s ( \sigma _{i_1} , \sigma _{i_2} , ... , \sigma _{i_s} ; t ) w \left(  4 \beta J m \sigma _{i_n} \right)  \bigr\} .
 \label{MF_dynam_sbody}
\end{eqnarray}
For an arbitrary $O(1)$ natural number $s$, the time development of $P_s$ under (\ref{MF_dynam_sbody}) is expressed as
\begin{equation}
 P_s ( \sigma _{i_1} , \sigma _{i_2} , ... , \sigma _{i_s} ; t ) = \prod _{ n=1} ^s p_1 ( \sigma _{i_n} , t ) ,
 \label{Ps1}
\end{equation}
if we ignore the $o \left( 1 / N \right)$ terms. Here, $p_1 ( \sigma ; t )$ is the solution of the following equation:
\begin{eqnarray}
 p_1 \left( \sigma , t+ \frac{1}{N} \right) & = & p_1 ( \sigma , t ) \nonumber \\
& & + \frac{1}{N} \bigl\{ p_1 ( -\sigma , t ) w \left( - 4 \beta J m \sigma \right) - p_1 ( \sigma , t ) w \left(  4 \beta J m \sigma \right)  \bigr\} ,
 \label{MF_dynam_1body}
\end{eqnarray}
This fact can be easily confirmed by substituting (\ref{Ps1}) and (\ref{MF_dynam_1body}) into (\ref{MF_dynam_sbody}).

Hence, when $N$ is large, the time development of every spin is independent of each other. The effect of the interaction appears only in the ``mean field,'' $m$, which is the effective external field comprising the mean value of the spins. Considering that $p_1$ can be expressed as 
\begin{equation}
 p_1 ( \sigma , t ) = \frac{1+m \sigma }{2} ,
 \label{p1_m}
\end{equation}
(\ref{MF_dynam_1body}) can also be expressed as
\begin{equation}
 m \left( t+ \frac{1}{N} \right) = m ( t ) + \frac{1}{N} \bigl\{ (1-m) w \left( - 4 \beta J m \right) - (1+m) w \left(  4 \beta J m \right)  \bigr\} .
 \label{MF_dynam_m}
\end{equation}
Taking the limit, $N \rightarrow \infty$, and substituting (\ref{transition1}), (\ref{MF_dynam_1body}) can be rewritten as
\begin{eqnarray}
\frac{dm}{dt} & = & (1-m) w \left( - 4 \beta J m \right) - (1+m) w \left(  4 \beta J m \right) \nonumber \\
& = & -m + \tanh \left( 2 \beta J m \right).
 \label{MF_dynam_m2}
\end{eqnarray}
In previous studies, the form of (\ref{MF_dynam_m2}) appears more frequently than (\ref{MF_dynam_1body})\cite{PHB89,SK68,CA99,OYCK00}.
 
\clearpage

\subsection{Calculation of spin correlation \label{2body} }

Substituting (\ref{transition2}) into (\ref{MF_dynam1}), the master equation in the finite-$N$ case becomes
\begin{eqnarray}
P_N \left( \left\{ \sigma \right\} ; t+ \frac{1}{N} \right) & = & P_N ( \left\{ \sigma \right\} ; t ) \nonumber \\
& & + \frac{1}{N} \sum _i \left[ P_N ( \left\{ F_i \sigma \right\} ; t ) \biggl\{ w \left( - 4 \beta J m \sigma _i  \right) \biggr. \right. \nonumber \\
& & - 4 \beta J \sigma _i \left( \frac{\sum _{j \neq i} \sigma _j }{N} -m \right) w' \left( - 4 \beta J m \sigma _i  \right) \nonumber \\
& & \left. + 8 \left( \beta J \right) ^2 \left( \frac{\sum _{j \neq i} \sigma _j }{N} -m \right) ^2 w'' \left( - 4 \beta J m \sigma _i  \right) \right\} \nonumber \\
& & - P_N ( \left\{ \sigma \right\} ; t ) \biggl\{ w \left(  4 \beta J m \sigma _i  \right) \biggr. \nonumber \\
& & + 4 \beta J \sigma _i \left( \frac{\sum _{j \neq i} \sigma _j }{N} -m \right) w' \left( 4 \beta J m \sigma _i  \right) \nonumber \\
& & \left. \left. + 8 \left( \beta J \right) ^2 \left( \frac{\sum _{j \neq i} \sigma _j }{N} -m \right) ^2 w'' \left( 4 \beta J m \sigma _i  \right) \right\} \right] ,
 \label{MF_dynam2_a}
\end{eqnarray}
Here, we consider the time development of this equation starting from the solution of the large-$N$ case discussed in section \ref{Ninf}.

Taking the difference between the function, $P_s$, and its zeroth approximation expressed in (\ref{Ps1}) as $\delta p_s$, we obtain the following expressions:
\numparts
\begin{eqnarray}
P_2 ( \sigma _i , \sigma _j ; t ) & \equiv & p_1(\sigma _i , t) p_1(\sigma _j , t) + \delta p_2 ( \sigma _i , \sigma _j ; t ) , \label{P2} \\
P_3 ( \sigma _i , \sigma _j , \sigma _k ; t ) & \equiv & p_1(\sigma _i , t) p_1(\sigma _j , t) p_1(\sigma _k , t) + \delta p_3 ( \sigma _i , \sigma _j , \sigma _k ; t ) , \label{P3} \\
P_4 ( \sigma _i , \sigma _j , \sigma _k , \sigma _l ; t ) & \equiv & p_1(\sigma _i , t) p_1(\sigma _j , t) p_1(\sigma _k , t) p_1(\sigma _l , t) + \delta p_4 ( \sigma _i , \sigma _j , \sigma _k , \sigma _l ; t ) , \label{P4}
\end{eqnarray}
\endnumparts
\begin{eqnarray}
\mathrm{where} \ \ \ \delta p_2 ( \sigma _i , \sigma _j ; t ) , \delta p_3 ( \sigma _i , \sigma _j , \sigma _k ; t ) , \delta p_4 ( \sigma _i , \sigma _j , \sigma _k , \sigma _l ; t ) = O \left( \frac{1}{N} \right) .
 \label{P2a}
\end{eqnarray}
In this paper, characters $i,j,k,$ and $l$ represent numbers different from each other unless there is a summation over them, such as $\sum _{i,j}$. Note that $p_1$ is not defined as the probability distribution of a single spin in the finite-size system, instead it is that in the thermodynamic limit discussed in section \ref{Ninf}. Specifically, $p_1$ itself is defined as the solution of (\ref{MF_dynam_1body}), and the modification of the finite-size system appears as $\delta p_2$, $\delta p_3$, and $\delta p_4$. We do not consider $P_s$ with $s \geq 5$, because it is not required for the calculation in this study.

When the summation is taken over all spin variables except $\sigma _i$ and $\sigma _j$, (\ref{MF_dynam2_a}) can be transformed into
\begin{eqnarray}
& & P_2 \left( \sigma _i , \sigma _j ; t+ \frac{1}{N} \right) - P_2 ( \sigma _i , \sigma _j ; t ) \nonumber \\
& = & \frac{1}{N} \left[  w \biggl( - 4 \beta J m \sigma _i  \right) P_2 ( - \sigma _i , \sigma _j ; t ) \biggr. \nonumber \\
& & -  \mathrm{Tr} _{ \left\{ \sigma _n \right\} _{n \neq i,j} } 4 \beta J \sigma _i \sum _k \left( \frac{ \sigma _k (1- \delta _{ki}) - m }{N} \right) w' \left( - 4 \beta J m \sigma _i  \right) P_N ( \left\{ F_i \sigma \right\} ; t ) \nonumber \\
& & + \mathrm{Tr} _{ \left\{ \sigma _n \right\} _{n \neq i,j} } 8 \left( \beta J \right) ^2 \sum _{k,l} \left( \frac{ \sigma _k (1- \delta _{ki}) - m }{N} \right) \left( \frac{ \sigma _l (1- \delta _{li}) - m }{N} \right) w'' \left( - 4 \beta J m \sigma _i  \right) P_N ( \left\{ F_i \sigma \right\} ; t ) \nonumber \\
& & + w \left( - 4 \beta J m \sigma _j  \right) P_2 ( \sigma _i , - \sigma _j ; t ) \nonumber \\
& & - \mathrm{Tr} _{ \left\{ \sigma _n \right\} _{n \neq i,j} } 4 \beta J \sigma _j \sum _k \left( \frac{ \sigma _k (1- \delta _{kj}) - m }{N} \right) w' \left( - 4 \beta J m \sigma _j  \right) P_N ( \left\{ F_j \sigma \right\} ; t ) \nonumber \\
& & + \mathrm{Tr} _{ \left\{ \sigma _n \right\} _{n \neq i,j} } 8 \left( \beta J \right) ^2 \sum _{k,l} \left( \frac{ \sigma _k (1- \delta _{kj}) - m }{N} \right) \left( \frac{ \sigma _l (1- \delta _{lj}) - m }{N} \right) w'' \left( - 4 \beta J m \sigma _j  \right) P_N ( \left\{ F_j \sigma \right\} ; t ) \nonumber \\
& & - w \left(  4 \beta J m \sigma _i  \right) P_2 (  \sigma _i , \sigma _j ; t ) \nonumber \\
& & - \mathrm{Tr} _{ \left\{ \sigma _n \right\} _{n \neq i,j} } 4 \beta J \sigma _i \sum _k \left( \frac{ \sigma _k (1- \delta _{ki}) - m }{N} \right) w' \left( 4 \beta J m \sigma _i  \right) P_N ( \left\{ \sigma \right\} ; t ) \nonumber \\ 
& & - \mathrm{Tr} _{ \left\{ \sigma _n \right\} _{n \neq i,j} } 8 \left( \beta J \right) ^2 \sum _{k,l} \left( \frac{ \sigma _k (1- \delta _{ki}) - m }{N} \right) \left( \frac{ \sigma _l (1- \delta _{li}) - m }{N} \right) w'' \left( 4 \beta J m \sigma _i  \right) P_N ( \left\{ \sigma \right\} ; t ) \nonumber \\
& & -w \left(  4 \beta J m \sigma _j  \right) P_2 (  \sigma _i , \sigma _j ; t ) \nonumber \\
& & - \mathrm{Tr} _{ \left\{ \sigma _n \right\} _{n \neq i,j} } 4 \beta J \sigma _j \sum _k \left( \frac{ \sigma _k (1- \delta _{kj}) - m }{N} \right) w' \left( 4 \beta J m \sigma _j  \right) P_N ( \left\{ \sigma \right\} ; t ) \nonumber \\
& & \left. - \mathrm{Tr} _{ \left\{ \sigma _n \right\} _{n \neq i,j} } 8 \left( \beta J \right) ^2 \sum _{k,l} \left( \frac{ \sigma _k (1- \delta _{kj}) - m }{N} \right) \left( \frac{ \sigma _l (1- \delta _{lj}) - m }{N} \right) w'' \left( 4 \beta J m \sigma _j  \right) P_N ( \left\{ \sigma \right\} ; t ) \right] .
 \label{MF_dynam2_b}
\end{eqnarray}

To simplify the right-hand side of (\ref{MF_dynam2_b}), we calculate the traces in the bracket. For example, the second term of the above equation is transformed into 
\begin{eqnarray}
& & \mathrm{Tr} _{ \left\{ \sigma _n \right\} _{n \neq i,j} } 4 \beta J \sigma _i \sum _k \left( \frac{ \sigma _k (1- \delta _{ki}) - m }{N} \right) w' \left( - 4 \beta J m \sigma _i  \right) P_N ( \left\{ F_i \sigma \right\} ; t ) \nonumber \\ 
& = & 4 \beta J \sigma _i \sum _{k \neq i,j} \sum _{ \sigma _k } \left( \frac{ \sigma _k - m }{N} \right) w' \left( - 4 \beta J m \sigma _i  \right) P_3 ( -\sigma _i , \sigma _j , \sigma _k ; t ) \nonumber \\
& & + 4 \beta J \sigma _i \left( \frac{ \sigma _j - 2m }{N} \right) w' \left( - 4 \beta J m \sigma _i  \right) P_2 ( -\sigma _i , \sigma _j ; t ) \nonumber \\
& = & 4 \beta J \sigma _i \sum _{ \sigma ' } \left( \sigma ' - m \right) w' \left( - 4 \beta J m \sigma _i  \right) \delta p_3 ( -\sigma _i , \sigma _j , \sigma ' ; t ) \nonumber \\
& & + 4 \beta J \sigma _i \left( \frac{ \sigma _j - 2m }{N} \right) w' \left( - 4 \beta J m \sigma _i  \right) p_1 ( -\sigma _i , t ) p_1( \sigma _j , t ) + o \left( \frac{1}{N} \right)
 \label{approx_p3}
\end{eqnarray}
Here, we used the following relations: 
\begin{eqnarray}
& & 4 \beta J \sigma _i \sum _{k \neq i,j} \sum _{ \sigma _k } \left( \frac{ \sigma _k - m }{N} \right) w' \left( - 4 \beta J m \sigma _i  \right) P_3( -\sigma _i , \sigma _j , \sigma _k ; t ) \nonumber \\
& = & 4 \beta J \sigma _i \sum _{k \neq i,j} \sum _{ \sigma _k } \left( \frac{ \sigma _k - m }{N} \right) w' \left( - 4 \beta J m \sigma _i  \right) p_1( -\sigma _i , t) p_1(\sigma _j , t) p_1(\sigma _k , t)  \nonumber \\
& & + 4 \beta J \sigma _i \sum _{k \neq i,j} \sum _{ \sigma _k } \left( \frac{ \sigma _k - m }{N} \right) w' \left( - 4 \beta J m \sigma _i  \right) \delta p_3( -\sigma _i , \sigma _j , \sigma _k ; t ) \nonumber \\
& = & 4 \beta J \sigma _i w' \left( - 4 \beta J m \sigma _i  \right) \sum _{ \sigma ' } \left( \sigma ' - m \right) \delta p_3( -\sigma _i , \sigma _j , \sigma ' ; t ) + o \left( \frac{1}{N} \right) , 
 \label{approx_p3a}
\end{eqnarray}
In the last line of (\ref{approx_p3a}), the equation, $\sum _{ \sigma _k } \sigma _k p_1(\sigma _k , t) = m $, is used. Similarly, the third term of (\ref{MF_dynam2_b}), which is proportional to $w'' \left( - 4 \beta J m \sigma _i  \right) $, can be transformed into
\begin{eqnarray}
& & \mathrm{Tr} _{ \left\{ \sigma _n \right\} _{n \neq i,j} } 8 \left( \beta J \right) ^2 \sum _{k,l} \left( \frac{ \sigma _k (1- \delta _{ki}) - m }{N} \right) \left( \frac{ \sigma _l (1- \delta _{li}) - m }{N} \right) w'' \left( - 4 \beta J m \sigma _i  \right) P_N ( \left\{ F_i \sigma \right\} ; t ) \nonumber \\ 
& = & 8 \left( \beta J \right) ^2 \sum _{k \neq i,j } \sum _{\sigma _k } \left( \frac{ \sigma _k - m }{N} \right) ^2 w'' \left( - 4 \beta J m \sigma _i  \right) p_1( -\sigma _i , t) p_1(\sigma _j , t) p_1(\sigma _k , t) \nonumber \\
& & + 8 \left( \beta J \right) ^2 \sum _{ \sigma ' , \sigma '' } \left( \sigma ' - m \right) \left( \sigma '' - m \right) w'' \left( - 4 \beta J m \sigma _i  \right) \delta p_4( -\sigma _i , \sigma _j , \sigma ' , \sigma '' ; t )  + o \left( \frac{1}{N} \right) \nonumber \\
& = & 8 \left( \beta J \right) ^2 \cdot \left( \frac{1 -m^2}{N} \right) \cdot w'' \left( - 4 \beta J m \sigma _i  \right) p_1( -\sigma _i , t) p_1(\sigma _j , t) \nonumber \\
& & + 8 \left( \beta J \right) ^2 \sum _{ \sigma ' , \sigma '' } \left( \sigma ' - m \right) \left( \sigma '' - m \right) w'' \left( - 4 \beta J m \sigma _i  \right) \delta p_4( -\sigma _i , \sigma _j , \sigma ' , \sigma '' ; t )  + o \left( \frac{1}{N} \right) 
 \label{approx_p4}
\end{eqnarray}
Here, we use the relation, 
\begin{equation}
\sum _{ \sigma _k } \left( \sigma _k - m \right) ^2 p_1(\sigma _k , t) = \sum _{ \sigma _k } \left( 1 + m^2 - 2m \sigma _k \right) p_1(\sigma _k , t) = (1 + m^2 ) - 2m \cdot m = 1 - m^2
\end{equation}
 Substituting these equations into (\ref{MF_dynam2_b}), we obtain
\begin{eqnarray}
& & P_2 \left( \sigma _i , \sigma _j ; t+ \frac{1}{N} \right) - P_2 ( \sigma _i , \sigma _j ; t ) \nonumber \\
& = & \frac{1}{N} \left[  w \left( - 4 \beta J m \sigma _i  \right) P_2 ( - \sigma _i , \sigma _j ; t ) -  4 \beta J \sigma _i \left( \frac{ \sigma _j - 2m }{N} \right) w' \left( - 4 \beta J m \sigma _i  \right) p_1 ( -\sigma _i , t) p_1( \sigma _j , t )  \right. \nonumber \\
& & - 4 \beta J \sigma _i w' \left( - 4 \beta J m \sigma _i  \right) \sum _{ \sigma ' } \left( \sigma ' - m \right) \delta p_3( -\sigma _i , \sigma _j , \sigma ' ; t ) \nonumber \\
& & + 8 \left( \beta J \right) ^2 \cdot \left( \frac{1 -m^2}{N} \right) \cdot w'' \left( - 4 \beta J m \sigma _i  \right) p_1( -\sigma _i , t) p_1(\sigma _j , t) \nonumber \\
& & + 8 \left( \beta J \right) ^2 \sum _{ \sigma ' , \sigma '' } \left( \sigma ' - m \right) \left( \sigma '' - m \right) w'' \left( - 4 \beta J m \sigma _i  \right) \delta p_4( -\sigma _i , \sigma _j , \sigma ' , \sigma '' ; t ) \nonumber \\
& & + w \left( - 4 \beta J m \sigma _j  \right) P_2 ( \sigma _i , - \sigma _j ; t ) - 4 \beta J \sigma _j \left( \frac{ \sigma _i - 2m }{N} \right) w' \left( - 4 \beta J m \sigma _j  \right) p_1 ( \sigma _i , t) p_1( -\sigma _j , t ) \nonumber \\
& & -  4 \beta J \sigma _j w' \left( - 4 \beta J m \sigma _j  \right) \sum _{ \sigma ' } \left( \sigma ' - m \right) \delta p_3( \sigma _i , -\sigma _j , \sigma ' ; t ) \nonumber \\
& & + 8 \left( \beta J \right) ^2 \cdot \left( \frac{1 -m^2}{N} \right) \cdot w'' \left( - 4 \beta J m \sigma _j  \right) p_1( \sigma _i , t) p_1( -\sigma _j , t) \nonumber \\
& & + 8 \left( \beta J \right) ^2 \sum _{ \sigma ' , \sigma '' } \left( \sigma ' - m \right) \left( \sigma '' - m \right) w'' \left( - 4 \beta J m \sigma _j  \right) \delta p_4( \sigma _i , -\sigma _j , \sigma ' , \sigma '' ; t ) \nonumber \\
& & - w \left(  4 \beta J m \sigma _i  \right) P_2 (  \sigma _i , \sigma _j ; t ) - 4 \beta J \sigma _i \left( \frac{ \sigma _j - 2m }{N} \right) w' \left( 4 \beta J m \sigma _i  \right) p_1 ( \sigma _i , t) p_1( \sigma _j , t ) \nonumber \\
& & - 4 \beta J \sigma _i w' \left( 4 \beta J m \sigma _i  \right) \sum _{ \sigma ' } \left( \sigma ' - m \right) \delta p_3( \sigma _i , \sigma _j , \sigma ' ; t ) \nonumber \\ 
& & - 8 \left( \beta J \right) ^2 \cdot \left( \frac{1 -m^2}{N} \right) \cdot w'' \left( 4 \beta J m \sigma _i  \right) p_1( \sigma _i , t) p_1( \sigma _j , t) \nonumber \\
& & - 8 \left( \beta J \right) ^2 \sum _{ \sigma ' , \sigma '' } \left( \sigma ' - m \right) \left( \sigma '' - m \right) w'' \left( 4 \beta J m \sigma _i  \right) \delta p_4( \sigma _i , \sigma _j , \sigma ' , \sigma '' ; t ) \nonumber \\
& & -w \left(  4 \beta J m \sigma _j  \right) P_2 (  \sigma _i , \sigma _j ; t ) - 4 \beta J \sigma _j \left( \frac{ \sigma _i - 2m }{N} \right) w' \left( 4 \beta J m \sigma _j  \right) p_1 ( \sigma _i , t) p_1( \sigma _j , t ) \nonumber \\
 & & - 4 \beta J \sigma _j w' \left( 4 \beta J m \sigma _j  \right) \sum _{ \sigma ' } \left( \sigma ' - m \right) \delta p_3( \sigma _i , \sigma _j , \sigma ' ; t ) \nonumber \\
& & - 8 \left( \beta J \right) ^2 \cdot \left( \frac{1 -m^2}{N} \right) \cdot w'' \left( 4 \beta J m \sigma _j  \right) p_1( \sigma _i , t) p_1( \sigma _j , t) \nonumber \\
& & \left. - 8 \left( \beta J \right) ^2 \sum _{ \sigma ' , \sigma '' } \left( \sigma ' - m \right) \left( \sigma '' - m \right) w'' \left( 4 \beta J m \sigma _j  \right) \delta p_4( \sigma _i , \sigma _j , \sigma ' , \sigma '' ; t ) \right] . 
 \label{MF_dynam2_c}
\end{eqnarray}
Extracting $O \left( 1/N ^2 \right)$ terms from (\ref{MF_dynam2_c}), the equation describing the time development of $\delta p_2$ is expressed as
\begin{eqnarray}
& & \delta p_2 \left( \sigma _i , \sigma _j , t+ \frac{1}{N} \right) - \delta p_2 ( \sigma _i , \sigma _j , t ) \nonumber \\
& = & - \frac{4 \beta J}{N^2} \left[ p _1 (\sigma _j ,t) p_1 (- \sigma _i ,t) \sigma _i ( \sigma _j - 2m ) w' (-4 \beta J m \sigma _i ) + p _1 (\sigma _j ,t) p_1 ( \sigma _i ,t) \sigma _i ( \sigma _j - 2m ) w' (4 \beta J m \sigma _i ) \right. \nonumber \\
& & \left. + p _1 (\sigma _i ,t) p_1 (- \sigma _j ,t) \sigma _j ( \sigma _i - 2m ) w' (-4 \beta J m \sigma _j ) + p _1 (\sigma _i ,t) p_1 ( \sigma _j ,t) \sigma _j ( \sigma _i - 2m ) w' (4 \beta J m \sigma _j ) \right] \nonumber \\
& & - \frac{1}{N^2} \bigl\{ p_1 ( - \sigma _i , t ) w \left( - 4 \beta J m \sigma _i  \right) - p_1 ( \sigma _i , t ) w \left(  4 \beta J m \sigma _i  \right) \bigr\} \nonumber \\
& & \cdot \bigl\{ p_1 ( - \sigma _j , t ) w \left( - 4 \beta J m \sigma _j  \right) - p_1 ( \sigma _j , t ) w \left(  4 \beta J m \sigma _j  \right) \bigr\} \nonumber \\
& & + \frac{1}{N} \bigl\{ \delta p_2 ( - \sigma _i , \sigma _j , t ) w \left( - 4 \beta J m \sigma _i  \right) + \delta p_2 ( \sigma _i , - \sigma _j , t ) w \left( - 4 \beta J m \sigma _j  \right) \bigr. \nonumber \\
& & - \delta p_2 ( \sigma _i , \sigma _j , t ) w \left(  4 \beta J m \sigma _i  \right) - \delta p_2 ( \sigma _i , \sigma _j , t ) w \left(  4 \beta J m \sigma _j  \right) \bigr\} \ \nonumber \\
& & - \frac{4 \beta J }{N} \left[ \sigma _i w' \left( - 4 \beta J m \sigma _i  \right) \sum _{ \sigma ' } \left( \sigma ' - m \right) \delta p_3( -\sigma _i , \sigma _j , \sigma ' ; t ) \right. \nonumber \\
& & + \sigma _j w' \left( - 4 \beta J m \sigma _j  \right) \sum _{ \sigma ' } \left( \sigma ' - m \right) \delta p_3( \sigma _i , -\sigma _j , \sigma ' ; t ) \nonumber \\
& & + \sigma _i w' \left( 4 \beta J m \sigma _i  \right) \sum _{ \sigma ' } \left( \sigma ' - m \right) \delta p_3( \sigma _i , \sigma _j , \sigma ' ; t ) \nonumber \\
& & \left. + \sigma _j w' \left( 4 \beta J m \sigma _j  \right)  \sum _{ \sigma ' } \left( \sigma ' - m \right) \delta p_3( \sigma _i , \sigma _j , \sigma ' ; t ) \right] \nonumber \\
& & + \left( \frac{ 8 \left( \beta J \right) ^2 \left( 1 -m^2 \right) }{N^2} \right) \cdot \left\{ w'' \left( - 4 \beta J m \sigma _i  \right) p_1( -\sigma _i , t) p_1(\sigma _j , t) + w'' \left( - 4 \beta J m \sigma _j  \right) p_1( \sigma _i , t) p_1( -\sigma _j , t) \right. \nonumber \\ 
& & \left. - w'' \left( 4 \beta J m \sigma _i  \right) p_1( \sigma _i , t) p_1( \sigma _j , t) - w'' \left( 4 \beta J m \sigma _j  \right) p_1( \sigma _i , t) p_1( \sigma _j , t)  \right\} \nonumber \\
& & + \frac{ 8 \left( \beta J \right) ^2 }{N} \left[ \sum _{ \sigma ' , \sigma '' } \left( \sigma ' - m \right) \left( \sigma '' - m \right) w'' \left( - 4 \beta J m \sigma _i  \right) \delta p_4( -\sigma _i , \sigma _j , \sigma ' , \sigma '' ; t ) \right. \nonumber \\
& & + \sum _{ \sigma ' , \sigma '' } \left( \sigma ' - m \right) \left( \sigma '' - m \right) w'' \left( - 4 \beta J m \sigma _j  \right) \delta p_4( \sigma _i , -\sigma _j , \sigma ' , \sigma '' ; t ) \nonumber \\
& & - \sum _{ \sigma ' , \sigma '' } \left( \sigma ' - m \right) \left( \sigma '' - m \right) w'' \left(  4 \beta J m \sigma _i  \right) \delta p_4( \sigma _i , \sigma _j , \sigma ' , \sigma '' ; t ) \nonumber \\
& & \left. - \sum _{ \sigma ' , \sigma '' } \left( \sigma ' - m \right) \left( \sigma '' - m \right) w'' \left(  4 \beta J m \sigma _j  \right) \delta p_4( \sigma _i , \sigma _j , \sigma ' , \sigma '' ; t ) \right] ,
\label{MF_dynam2body_a}
\end{eqnarray}
Here, the first term of (\ref{MF_dynam2body_a}) can be transformed into the following simpler form:
\begin{eqnarray}
&  & - \frac{4 \beta J}{N^2} \left[ p _1 (\sigma _j ,t) p_1 (- \sigma _i ,t) \sigma _i ( \sigma _j - 2m ) w' (-4 \beta J m \sigma _i ) + p _1 (\sigma _j ,t) p_1 ( \sigma _i ,t) \sigma _i ( \sigma _j - 2m ) w' (4 \beta J m \sigma _i ) \right. \nonumber \\
& & \left. + p _1 (\sigma _i ,t) p_1 (- \sigma _j ,t) \sigma _j ( \sigma _i - 2m ) w' (-4 \beta J m \sigma _j ) + p _1 (\sigma _i ,t) p_1 ( \sigma _j ,t) \sigma _j ( \sigma _i - 2m ) w' (4 \beta J m \sigma _j ) \right] \nonumber \\
& = & + \frac{4 \beta J \sigma _i }{N^2} \left[ w' ( 4 \beta J m \sigma _i ) p _1 ( \sigma _i ,t) + w' (-4 \beta J m \sigma _i ) p _1 ( -\sigma _i ,t) \right] \left\{ \frac{m}{2} - \frac{1-2m^2}{2} \sigma _j \right\} \nonumber \\
& & + \frac{4 \beta J \sigma _j }{N^2} \left[ w' ( 4 \beta J m \sigma _j ) p _1 ( \sigma _j ,t) + w' (-4 \beta J m \sigma _j ) p _1 ( -\sigma _j ,t) \right] \left\{ \frac{m}{2} - \frac{1-2m^2}{2} \sigma _i \right\} \nonumber \\
& = & \frac{4 \beta J }{N^2} \left[ w' ( 4 \beta J m ) p _1 ( 1,t ) + w' (-4 \beta J m ) p _1 ( -1,t ) \right] \left\{ \frac{m}{2} \left( \sigma _i + \sigma _j \right) - (1- 2m^2 ) \sigma _i \sigma _j \right\} 
\label{MF_dynam2body_a1}
\end{eqnarray}

Subsequently, we introduce $O \left( 1/N \right)$ parameters $\delta m$ and $v$, and let
\numparts
\begin{eqnarray}
\left< \sigma _i \right> & \equiv & m + \delta m , \\
\left< \delta \sigma _i \delta \sigma _j \right> & \equiv & v , 
 \label{mv1}
\end{eqnarray}
\endnumparts
\begin{equation}
\mathrm{where} \ \ \ \delta \sigma _i \equiv \sigma _i - \left< \sigma _i \right> .
\end{equation}
Assuming that $p_2, p_3$, and $p_4$ present symmetry under the permutation of variables, these functions are expressed as 
\numparts
\begin{eqnarray}
\delta p_2 ( \sigma _i , \sigma _j ; t ) & = & \frac{\delta m}{4} \left( \sigma _i + \sigma _j \right) + \frac{v + 2m \delta m }{4} \sigma _i \sigma _j \label{p2} \\
\delta p_3 ( \sigma _i , \sigma _j , \sigma _k ; t ) & = & \frac{\delta m}{8} \left( \sigma _i + \sigma _j + \sigma _k \right)  + \frac{v + 2m \delta m }{8} \left( \sigma _i \sigma _j + \sigma _j \sigma _k + \sigma _k \sigma _i \right) + \alpha _3 \sigma _i \sigma _j \sigma _k \label{p3} \\
\delta p_4 ( \sigma _i , \sigma _j , \sigma _k , \sigma _l ; t ) & = & \frac{\delta m}{16} \left( \sigma _i + \sigma _j + \sigma _k + \sigma _l \right)  + \frac{v + 2m \delta m }{16} \left( \sigma _i \sigma _j + \sigma _i \sigma _k + \sigma _i \sigma _l + \sigma _j \sigma _k + \sigma _j \sigma _l + \sigma _k \sigma _l \right) \nonumber \\
& & + \frac{\alpha _3}{2} \left( \sigma _i \sigma _j \sigma _k + \sigma _i \sigma _j \sigma _l + \sigma _i \sigma _k \sigma _l + \sigma _j \sigma _k \sigma _l \right) + \alpha _4 \sigma _i \sigma _j \sigma _k \sigma _l , \label{p4}
\end{eqnarray}
\endnumparts
Note that the two-body correlation, $v$, is closely related to the fluctuation of magnetization as follows:
\begin{eqnarray}
\left< \delta M^2 \right> & \equiv & \left< M^2 \right> - \left< M \right> ^2 = N (N-1) \left< \sigma _i \sigma _j \right> + N \left< \sigma _i ^2 \right> - N ^2 \left< \sigma _i \right> ^2 \nonumber \\
& = & N (N-1) \left\{ v + \left( m + \delta m \right) ^2 \right\} + N - N^2 \left( m + \delta m \right) ^2 + O(1) \nonumber \\
& = & N \left( Nv + 1-m^2 \right) + O(1) .
\label{deltaM}
\end{eqnarray}
Based on (\ref{MF_dynam2body_a}), the information on $\delta p_3$ and $\delta p_4$ is necessary to calculate the time development of $\delta p_2$. This reflects the problem of BBGKY hierarchy\cite{CDFR14}. In this study, an assumption is introduced to deal with this problem by expressing $\alpha _3$ using $m, \delta m$, and $v$. Specifically, we assume that the probability density of magnetization $M/N = \left( \sum _i \sigma _i \right) /N $ can be approximated in a Gaussian form. Hence, the third-order cumulant of this property is approximately zero.
\begin{equation}
\frac{1}{N^3} \left( \left< \sum _{i,j,k} \sigma _i \sigma _j \sigma _k \right> - 3 \left< \sum _{i,j} \sigma _i \sigma _j \right> \left< \sum _i \sigma _i \right> + 2 \left< \sum _i \sigma _i \right> ^3 \right) = 0 . 
 \label{mv2a}
\end{equation}
In several previous studies, similar assumption is adopted to delete higher-order cumulants \cite{LT10,PT18}. In the case of an equilibrium state, (\ref{mv2a}) can be confirmed by calculating the second-order derivative of $\left< M \right>$. Specifically, considering that
\begin{equation}
 \frac{1}{\beta ^2} \left. \frac{\partial ^2 \left< M \right> }{\partial h ^2} \right| _{h \rightarrow 0} = \left< M ^3 \right> - 3 \left< M ^2 \right> \left< M \right> + 2 \left< M \right> ^3 = O(N) , 
\end{equation}
the left-hand side of (\ref{mv2a}) is the infinitesimal of $O(1/N^2)$, which is ignored in this study. 

Calculating the left-hand side of (\ref{mv2a}), we obtain
\begin{eqnarray}
& & \frac{(N-1)(N-2) }{N^2} \left< \sigma _i \sigma _j \sigma _k \right> + \frac{3N-2 }{N^2} \left< \sigma _i \right> - 3 \frac{ (N-1) \left< \sigma _i \sigma _j \right> + 1 }{N} \left< \sigma _i \right> + 2 \left< \sigma _i \right> ^3 \nonumber \\
& = & \left( 1 - \frac{3}{N} \right) \left< \sigma _i \sigma _j \sigma _k \right> + \frac{3}{N} m - 3m \left( m^2 + v + 2m \delta m + \frac{1-m^2 }{N} \right) \nonumber \\
& & - 3m^2 \delta m + 2m^3 + 6m^2 \delta m + O \left( \frac{1}{N^2} \right) \nonumber \\
& = & 8 \alpha _3 -3mv - 3m^2 \delta m  + O \left( \frac{1}{N^2} \right) = O \left( \frac{1}{N^2} \right) . 
 \label{mv2b}
\end{eqnarray}
 Hence, ignoring $O(1/N^2 )$ terms, $\alpha _3 $ is expressed as 
\begin{equation}
\alpha _3 = \frac{3m \left( v + m \delta m \right) }{8} . 
 \label{alpha3}
\end{equation} 
Using (\ref{p2})--(\ref{p4}), the following relations are confirmed:
\numparts
\begin{eqnarray}
 \sum _{\sigma ' } ( \sigma ' - m ) \delta p_3 ( \sigma _i , \sigma _j , \sigma ' ; t ) & = & \frac{\delta m}{4} + \frac{v+ m \delta m}{4} \left( \sigma _i + \sigma _j \right) + \alpha ' _3 \sigma _i \sigma _j , \\
 \sum _{\sigma ' , \sigma '' } ( \sigma ' - m ) ( \sigma '' - m ) \delta p_4 ( \sigma _i , \sigma _j , \sigma ' , \sigma '' ; t ) & = & \frac{v}{4} + \left( \frac{m(-2v - 3m \delta m ) }{4} + 2 \alpha _3 \right) \left( \sigma _i + \sigma _j \right) + \alpha ' _4 \sigma _i \sigma _j \nonumber \\
& = &  \frac{v}{4} + \frac{mv }{4} \left( \sigma _i + \sigma _j \right) + \alpha ' _4 \sigma _i \sigma _j .
\end{eqnarray}
\endnumparts
\begin{eqnarray}
\mathrm{where} \ \ \left\{
\begin{array}{ccc}
\alpha ' _3 & = & 2 \alpha _3  -  \frac{m \left( v + 2m \delta m \right)}{4}  \\
\alpha ' _4 & = & 4 \alpha _4 - 4\alpha _3 m + \frac{m ^2 \left( v + 2m \delta m \right)}{4}  \\
\end{array} 
\right.
\end{eqnarray}
We do not calculate the concrete forms of $\alpha ' _3$ and $\alpha ' _4$ because they are unnecessary, which is subsequently explained. Using these coefficients and reflecting (\ref{MF_dynam2body_a1}), (\ref{MF_dynam2body_a}) is transformed into
\begin{eqnarray}
& & \delta p_2 \left( \sigma _i , \sigma _j , t+ \frac{1}{N} \right) - \delta p_2 ( \sigma _i , \sigma _j , t ) \nonumber \\
& = & \frac{4 \beta J }{N^2} \left[ w' ( 4 \beta J m ) p _1 ( 1,t ) + w' (-4 \beta J m ) p _1 ( -1,t ) \right] \left\{ \frac{m}{2} \left( \sigma _i + \sigma _j \right) - (1- 2m^2 ) \sigma _i \sigma _j \right\} \nonumber \\
& & - \frac{1}{N^2} \bigl\{ p_1 ( - 1 , t ) w \left( - 4 \beta J m \right) - p_1 ( 1 , t ) w \left(  4 \beta J m \right) \bigr\} ^2 \sigma _i \sigma _j \nonumber \\
& & - \frac{\delta m }{4N} \left( \sigma _i + \sigma _j  \right) \bigl\{ w \left( - 4 \beta J m \right) + w \left( 4 \beta J m \right) \bigr\} + \frac{\delta m }{2N} \sigma _i \sigma _j \bigl\{ w \left( - 4 \beta J m \right) - w \left( 4 \beta J m \right) \bigr\} \nonumber \\
& & - \frac{v + 2m \delta m}{2N} \sigma _i \sigma _j \bigl\{ w \left( - 4 \beta J m \right) + w \left( 4 \beta J m \right) \bigr\} \nonumber \\
& & - \frac{4 \beta J }{N} \left[ \frac{\delta m}{4} \left( \sigma _i + \sigma _j  \right) \bigl\{ w' \left( - 4 \beta J m \right) + w' \left( 4 \beta J m \right) \bigr\} + 2 \alpha ' _3 \sigma _i \sigma _j \bigl\{ w' \left( 4 \beta J m \right) - w' \left( - 4 \beta J m \right) \bigr\} \right. \nonumber \\
& & \left. + \frac{v + m\delta m}{4} \left[ 2 \sigma _i \sigma _j \bigl\{ w' \left( 4 \beta J m \right) + w' \left( - 4 \beta J m \right) \bigr\} + \left( \sigma _i + \sigma _j  \right) \bigl\{ - w' \left( - 4 \beta J m \right) + w' \left( 4 \beta J m \right) \bigr\} \right] \right] \nonumber \\
& & - \left( \frac{ 2 \left( \beta J \right) ^2 \left( 1 -m^2 \right) }{N^2} \right) \cdot \left( \sigma _i + \sigma _j + 2m \sigma _i \sigma _j \right) \nonumber \\
& & \cdot \left[ w'' \left( 4 \beta J m \right) - w'' \left( -4 \beta J m \right) + m \left\{ w'' \left( 4 \beta J m \right) + w'' \left( -4 \beta J m \right) \right\} \right] \nonumber \\
& & + \frac{ 8 \left( \beta J \right) ^2 }{N} \left[ -\frac{v}{4} \left( \sigma _i + \sigma _j \right) \left\{ w'' \left( 4 \beta J m \right) - w'' \left( -4 \beta J m \right) \right\} - 2 \alpha ' _4 \sigma _i \sigma _j \left\{ w'' \left( 4 \beta J m \right) + w'' \left( -4 \beta J m \right) \right\} \right. \nonumber \\
& & \left. - \frac{mv}{4} \left( \sigma _i + \sigma _j \right) \left\{ w'' \left( 4 \beta J m \right) + w'' \left( -4 \beta J m \right) \right\} - \frac{mv}{2} \sigma _i \sigma _j \left\{ w'' \left( 4 \beta J m \right) - w'' \left( -4 \beta J m \right) \right\} \right] 
\label{MF_dynam2body_2}
\end{eqnarray}
Comparing (\ref{p2}) and (\ref{MF_dynam2body_2}), we obtain the following equations:
\begin{eqnarray}
& & \frac{1}{4} \left\{ \delta m \left( t+ \frac{1}{N} \right) - \delta m ( t ) \right\} \nonumber \\
& = & + \frac{4 \beta J }{N^2} \left[ w' ( 4 \beta J m ) p _1 ( 1,t ) + w' (-4 \beta J m ) p _1 ( -1,t ) \right] \frac{m}{2} - \frac{\delta m }{4N} \bigl\{ w \left( - 4 \beta J m \right) + w \left( 4 \beta J m \right) \bigr\} \nonumber \\
& & - \frac{4 \beta J }{N} \left[ \frac{\delta m}{4} \bigl\{ w' \left( - 4 \beta J m \right) + w' \left( 4 \beta J m \right) \bigr\} + \frac{v + m\delta m}{4} \bigl\{ - w' \left( - 4 \beta J m \right) + w' \left( 4 \beta J m \right) \bigr\} \right]  \nonumber \\
\nonumber \\
& & - \left( \frac{ 2 \left( \beta J \right) ^2 \left( 1 -m^2 \right) }{N^2} \right) \cdot \left[ w'' \left( 4 \beta J m \right) - w'' \left( -4 \beta J m \right) + m \left\{ w'' \left( 4 \beta J m \right) + w'' \left( -4 \beta J m \right) \right\} \right] \nonumber \\
& & + \frac{ 8 \left( \beta J \right) ^2 }{N} \left[ -\frac{v}{4} \left\{ w'' \left( 4 \beta J m \right) - w'' \left( -4 \beta J m \right) \right\} - \frac{mv}{4} \left\{ w'' \left( 4 \beta J m \right) + w'' \left( -4 \beta J m \right) \right\} \right] \nonumber \\
\label{MF_dynam2body_3a}
\end{eqnarray}
\begin{eqnarray}
& & \frac{1}{4} \left[ \left\{ v \left( t+ \frac{1}{N} \right) + 2 m \left( t+ \frac{1}{N} \right) \delta m \left( t+ \frac{1}{N} \right) \right\} - \left\{ v(t) + 2m(t) \delta m (t) \right\} \right] \nonumber \\
& = & - \frac{4 \beta J }{N^2} \left[ w' ( 4 \beta J m ) p _1 ( 1,t ) + w' (-4 \beta J m ) p _1 ( -1,t ) \right] (1- 2m^2 ) \nonumber \\
& & - \frac{1}{N^2} \bigl\{ p_1 ( - 1 , t ) w \left( - 4 \beta J m \right) - p_1 ( 1 , t ) w \left(  4 \beta J m \right) \bigr\} ^2 \nonumber \\
& & + \frac{\delta m }{2N} \bigl\{ w \left( - 4 \beta J m \right) - w \left( 4 \beta J m \right) \bigr\} - \frac{v + 2m \delta m}{2N} \bigl\{ w \left( - 4 \beta J m \right) + w \left( 4 \beta J m \right) \bigr\} \nonumber \\
& & - \frac{4 \beta J }{N} \left[ 2 \alpha ' _3 \bigl\{ w' \left( 4 \beta J m \right) - w' \left( - 4 \beta J m \right) \bigr\} + \frac{v + m\delta m}{2} \bigl\{ w' \left( 4 \beta J m \right) + w' \left( - 4 \beta J m \right) \bigr\} \right] \nonumber \\
& & - \left( \frac{ 4m \left( \beta J \right) ^2 \left( 1 -m^2 \right) }{N^2} \right) \cdot \left[ w'' \left( 4 \beta J m \right) - w'' \left( -4 \beta J m \right) + m \left\{ w'' \left( 4 \beta J m \right) + w'' \left( -4 \beta J m \right) \right\} \right] \nonumber \\
& & + \frac{ 8 \left( \beta J \right) ^2 }{N} \left[ - 2 \alpha ' _4 \left\{ w'' \left( 4 \beta J m \right) + w'' \left( -4 \beta J m \right) \right\} - \frac{mv}{2} \left\{ w'' \left( 4 \beta J m \right) - w'' \left( -4 \beta J m \right) \right\} \right] . 
\label{MF_dynam2body_3b}
\end{eqnarray}

Rearranging the above equations and taking the limit that $N$ is sufficiently large, (\ref{MF_dynam2body_3a}) and (\ref{MF_dynam2body_3b}) can be transformed as the differential equations,
\begin{eqnarray}
& & \frac{ d \left( \delta m \right)}{dt} \nonumber \\
& = & \frac{8 \beta J }{N} \left[ w' ( 4 \beta J m ) p _1 ( 1,t ) + w' (-4 \beta J m ) p _1 ( -1,t ) \right] m - \delta m \bigl\{ w \left( - 4 \beta J m \right) + w \left( 4 \beta J m \right) \bigr\} \nonumber \\
& & - 4 \beta J  \left[ \delta m \bigl\{ w' \left( - 4 \beta J m \right) + w' \left( 4 \beta J m \right) \bigr\} + \left( v + m \delta m \right) \bigl\{ - w' \left( - 4 \beta J m \right) + w' \left( 4 \beta J m \right) \bigr\} \right] \nonumber \\
& & - \left( \frac{ 8 \left( \beta J \right) ^2 \left( 1 -m^2 \right) }{N} \right) \cdot \left[ w'' \left( 4 \beta J m \right) - w'' \left( -4 \beta J m \right) + m \left\{ w'' \left( 4 \beta J m \right) + w'' \left( -4 \beta J m \right) \right\} \right] \nonumber \\
& & + 8 \left( \beta J \right) ^2 \left[ -v \left\{ w'' \left( 4 \beta J m \right) - w'' \left( -4 \beta J m \right) \right\} - mv \left\{ w'' \left( 4 \beta J m \right) + w'' \left( -4 \beta J m \right) \right\} \right] , 
\label{MF_dynam2body_3a2}
\end{eqnarray}
\begin{eqnarray}
& & \frac{dv}{dt} \nonumber \\
& = & - \frac{16 \beta J }{N} \left[ w' ( 4 \beta J m ) p _1 ( 1,t ) + w' (-4 \beta J m ) p _1 ( -1,t ) \right] \left( 1 - m^2 \right) \nonumber \\
& & - \frac{4}{N} \bigl\{ p_1 ( - 1 , t ) w \left( - 4 \beta J m \right) - p_1 ( 1 , t ) w \left(  4 \beta J m \right) \bigr\} ^2 - 2 v \bigl\{ w \left( - 4 \beta J m \right) + w \left( 4 \beta J m \right) \bigr\} \nonumber \\
& & - 8 \beta J \left[ \left( 4 \alpha ' _3 - m \left( v + m \delta m \right)  \right) \bigl\{ w' \left( 4 \beta J m \right) - w' \left( - 4 \beta J m \right) \bigr\} + v \bigl\{ w' \left( 4 \beta J m \right) + w' \left( - 4 \beta J m \right) \bigr\} \right] \nonumber \\
& & + 16 \left( \beta J \right) ^2 \left( -4 \alpha ' _4 + m ^2 v \right) \left\{ w'' \left( 4 \beta J m \right) + w'' \left( -4 \beta J m \right) \right\}  . 
\label{MF_dynam2body_3d}
\end{eqnarray}
Note that the above calculations does not use the concrete form of $w(x)$ of the Glauber dynamics expressed in (\ref{transition1}). However, if $w(x)$ or its derivatives have singular points, these approaches, which are based on the Taylor expansion of (\ref{transition2}), break down. Hence, we cannot apply (\ref{MF_dynam2body_3a2}) and (\ref{MF_dynam2body_3d}) in the case of the Metropolis method, $w(x) = \min \left( 1, e^{-x} \right) $.

Substituting (\ref{transition1}) into (\ref{MF_dynam2body_3a2}) and (\ref{MF_dynam2body_3d}), we obtain
\begin{equation}
\frac{ d \left( \delta m \right)}{dt} = \frac{2 \beta J }{ \cosh ^2 \left( 2 \beta Jm \right) } \cdot \left( \delta m - \frac{m}{N} \right) - \delta m - \left( \frac{ 4 \left( \beta J \right) ^2 \sinh \left( 2 \beta Jm \right) }{\cosh ^3 \left( 2 \beta Jm \right) } \right) \cdot \left( v + \frac{1-m^2}{N} \right) ,
\label{MF_dynam2body_4a2}
\end{equation}
\begin{equation}
\frac{dv}{dt} = \frac{4 \beta J }{\cosh ^2 \left( 2 \beta Jm \right) } \cdot \left( v + \frac{1-m^2}{N} \right) - \frac{1}{N} \bigl\{ m - \tanh \left( 2 \beta J m \right) \bigr\} ^2 - 2 v . 
\label{MF_dynam2body_4d}
\end{equation}

Note that these equations do not contain $\alpha ' _3$ and $\alpha ' _4$ because $w'(x) - w'(-x) = w''(x) + w''(-x) =0$ under the Glauber dynamics.
We briefly discuss the fluctuation of magnetization in the case of $T \geq T_c$. In this case, when a sufficiently long time has passed, the magnetization is approximately zero. Hence, we can transform (\ref{MF_dynam2body_4d}) into a simple form by substituting $m \simeq 0$: 
\begin{equation}
\frac{dv}{dt} = \frac{4 \beta J }{N} - 2 \left( 1 - 2 \beta J \right) v . 
\label{MF_dynam2body_3d_Tc}
\end{equation}
Based on the above equation, $v$ approaches its equilibrium value exponentially when $T > T_c$. However, if $T=T_c (= 2J)$, coefficient $1-2 \beta J$ in this equation becomes zero, and $v$ increases with time linearly as follows:
\begin{equation}
v = \frac{2 }{N}  t + \mathrm{const} . 
 \label{MF_v_Tc}
\end{equation}
This reflects that the fluctuation of the magnetization in the equilibrium state (precisely, that multiplied by the coefficient, $1/N$, to regulate the order), $\left< \delta M^2 \right> /N$, diverges at this temperature. Considering that (\ref{MF_dynam2body_4a2}) and (\ref{MF_dynam2body_4d}) are derived by assuming that this fluctuation is small, their accuracy is expected to become worse with time. However, when $N$ is large, the time range in which we can use these equations is lengthened because $v$ is inversely proportional to $N$. Time development at the critical temperature was already discussed by Anteneodo \textit{et al.}, and they found a linear increase in the fluctuation of magnetization similar to (\ref{MF_v_Tc}) \cite{AFC10}. 
Note that the discussion on $\delta m$ at $T = T_c$ is more complex than that of $v$, because $v$ increases with time. Under this effect, we cannot conclude that the last term of the right-hand side of (\ref{MF_dynam2body_4a2}) converges to zero in the large-$t$ limit even though $m \simeq 0$. 

\begin{figure}[hbp!]
\begin{center}
\includegraphics[width = 15.0cm]{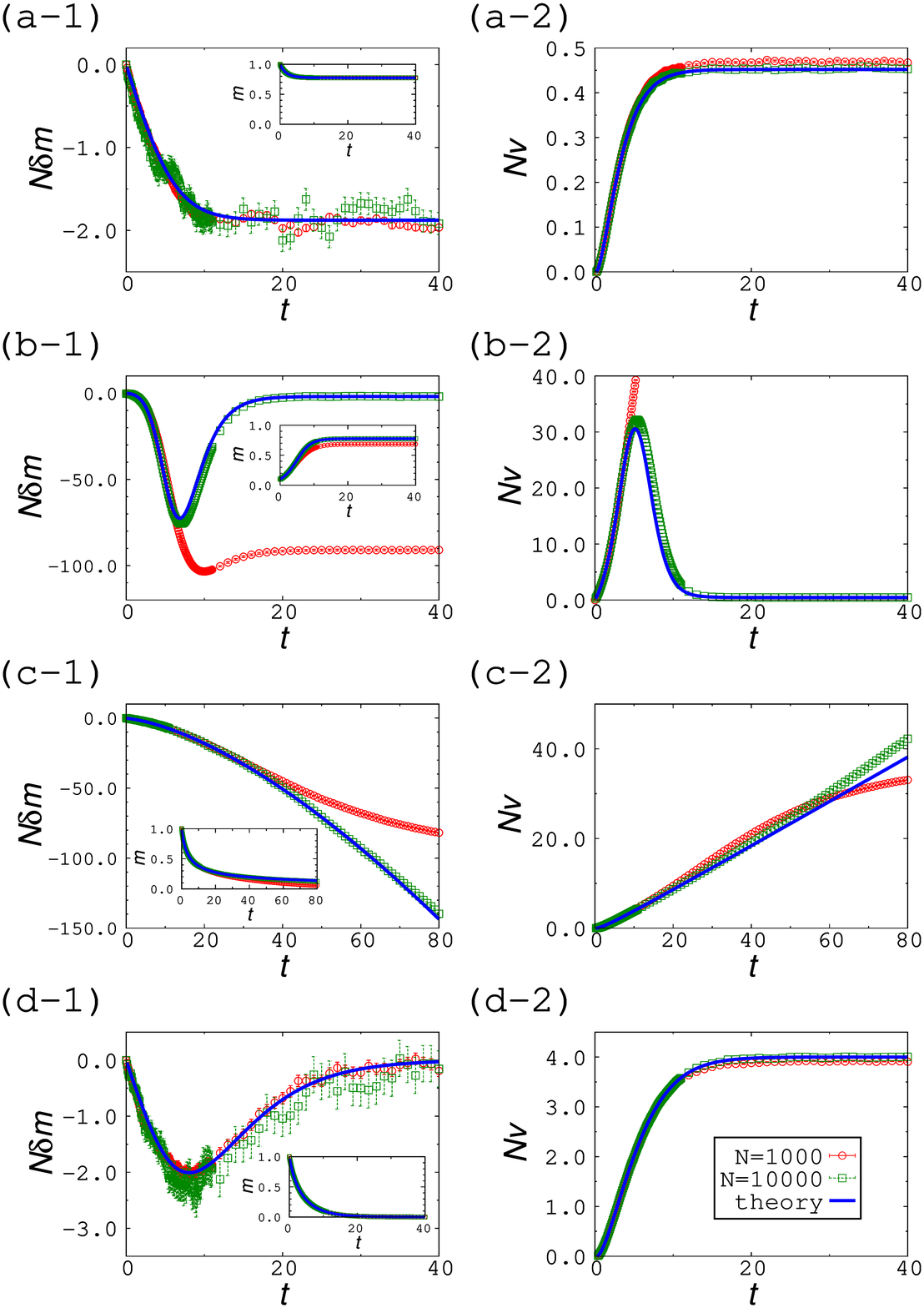} 
\end{center}
\caption{(Color online) Time development of (a-1)$N\delta m$ and (a-2)$Nv$ at $(T,m_0) = (1.5(<T_c), 1.0)$, (b-1)$N\delta m$ and (b-2)$Nv$ at $(T,m_0) = (1.5, 0.1)$, (c-1)$N\delta m$ and (c-2)$Nv$ at $(T,m_0) = (2.0(=T_c), 1.0)$, and (d-1)$N\delta m$ and (d-2)$Nv$ at $(T,m_0) = (2.5(>T_c), 1.0)$, respectively. Red circle and green square points present results of MCMC simulations at $N = 1000$ and $N = 10000$, respectively, and blue curves are numerical solutions of (\ref{MF_dynam2body_3a2}) and (\ref{MF_dynam2body_3d}). Inset of each $N \delta m$ graph shows comparison of magnetization calculated from simulations and numerical solution of (\ref{MF_dynam_m2}).  }
\label{dmvsim}
\end{figure}

To investigate the accuracy of the above method, we calculate the numerical solutions of (\ref{MF_dynam2body_4a2}) and (\ref{MF_dynam2body_4d}) using the fourth-order Runge--Kutta method with time interval $\delta t = 1.0 \times 10^{-3}$ and compare them with the results of actual MCMC simulations. In these simulations, averages over 480,000 independent trials are taken for each property, and $J$ is fixed as $J = 1$. Note that in the actual simulations, $\delta m$ is defined as the difference between the calculated magnetization and the theoretical value of $m$ in the thermodynamic limit obtained using (\ref{MF_dynam_m2}) with time interval $\delta t = 1/N$. In the initial state of the simulations, each spin has an independent probability distribution with $m = m_0$, i.e., $\delta m = v = 0$. The results for the cases of $(T,m_0) = (1.5(<T_c) , 1.0)$, $(1.5, 0.1)$, $(2.0(=T_c), 1.0)$, and $(2.5(>T_c), 1.0)$ are shown in Figures \ref{dmvsim} (a)--(d), respectively. Here, we plot $N \delta m$ and $Nv$ in the left and right graphs, respectively. We also draw a graph comparing the magnetization calculated from the simulations and the numerical solutions of (\ref{MF_dynam_m2}) as insets in the left graphs.
The above figure shows that the solutions of (\ref{MF_dynam2body_3a2}) and (\ref{MF_dynam2body_3d}) fit the actual simulation results well when $(T,m_0) = (1.5 , 1.0)$ and $(2.5 , 1.0)$(graphs (a) and (d)). However, their differences become large in the other two cases(graphs (b) and (c)), particularly when $N$ is small. The cause of this difference when $T=2.0 (= T_c)$ is already discussed in above. When $(T,m_0) = (1.5 , 0.1)$, this difference is considered to originate from the reversal of the magnetization. Considering that the height of the energy barrier when the magnetization changes from $m_0$ to 0 is broadly evaluated as $NJm_0 ^2$, magnetization reversal frequently occurs when $N$ and $m_0$ are small. In fact, the inset in Figure \ref{dmvsim}(b-1) shows that the magnetization at $N = 1000$ is more different from the solution of (\ref{MF_dynam_m2}) than in the other cases. This implies that the zeroth approximation discussed in section \ref{Ninf} itself is broken by the cases in which magnetization reversal occurs. 

Briefly, the approximation in this study breaks down when the fluctuation of the system is large because of critical phenomena or magnetization reversal occurrence.

\clearpage

\subsection{Calculation of autocorrelation of magnetization \label{auto_corr} }

In this section, calculation of the autocorrelation of magnetization is presented. Its derivation is similar to that provided in the previous section, except that the conditional probability distribution is used instead of $P_N$. We express the conditional probability of the spin configuration, $\left\{ \sigma \right\}$, at time $t(>t_0)$ under the condition, $\left\{ \sigma \right\} = \left\{ \sigma ' \right\}$, at time $t_0$ as $\tilde{P} _{N} ( \left\{ \sigma \right\} ; t | \left\{ \sigma ' \right\} , t_0 )$, and define the following:
\begin{eqnarray}
G_{N} \left( \left\{ \sigma \right\} ; t | t_0 \right) & \equiv & \mathrm{Tr} _{ \left\{ \sigma ' \right\} } \sum _{i } \frac{1}{N} \sigma ' _{i} P_N ( \left\{ \sigma ' \right\} , t_0 ) \tilde{P} _{N} ( \left\{ \sigma \right\} ; t | \left\{ \sigma ' \right\} , t_0 ) , \\
G_{s} \left( \sigma _{i_1} , \sigma _{i_2} , ... , \sigma _{i_s} ; t | t_0 \right) & \equiv & \mathrm{Tr} _{ \left\{ \sigma _n \right\} _{n \neq i_1 , i_2 , ... , i_s } } G_{N} \left( \left\{ \sigma \right\} ; t | t_0 \right) .
 \label{Gi}
\end{eqnarray}
To obtain the autocorrelation, first the time development of $G_2$ needs to be derived, for which the following relation is used:
\begin{eqnarray}
\frac{1}{N^2} \left< M(t_0) M(t) \right> = \frac{1}{N} \sum _i \mathrm{Tr} _{\left\{ \sigma \right\} } \sigma _i G_N \left( \left\{ \sigma \right\} ; t | t_0 \right) = \sum _{ \sigma , \sigma ' } \sigma G_{2} \left( \sigma  , \sigma ' ; t | t_0 \right) .  
\label{autoC}
\end{eqnarray}

Considering that $\tilde{P} _{N} ( \left\{ \sigma \right\} ; t | \left\{ \sigma ' \right\} , t_0 )$ and $G_{N} ( t_0 | \left\{ \sigma \right\} ; t )$ obey the master equation in (\ref{MF_dynam2_a}), similar to $P_N ( \left\{ \sigma \right\} ; t )$, the time development of $G_{2} \left( \sigma _i , \sigma _j ; t | t_0 \right) $ can be written as
\begin{eqnarray}
& & G_{2} \left( \sigma _i , \sigma _j ; t+ \left. \frac{1}{N} \right| t_0 \right) - G_{2} ( \sigma _i , \sigma _j ; t | t_0 ) \nonumber \\
& = & \frac{1}{N} \biggl[  w \left( - 4 \beta J m(t) \sigma _i  \right) G_{2} ( - \sigma _i , \sigma _j ; t | t_0 ) \biggr. \nonumber \\ 
& & -  \mathrm{Tr} _{ \left\{ \sigma _n \right\} _{n \neq i,j} } 4 \beta J \sigma _i \sum _k \left( \frac{ \sigma _k (1- \delta _{ki}) - m(t) }{N} \right) w' \left( - 4 \beta J m(t) \sigma _i  \right) G_{N} ( \left\{ F_i \sigma \right\} ; t | t_0 ) \nonumber \\
& & + \mathrm{Tr} _{ \left\{ \sigma _n \right\} _{n \neq i,j} } 8 \left( \beta J \right) ^2 \sum _{k,l} \nonumber \\ 
& & \cdot \left( \frac{ \sigma _k (1- \delta _{ki}) - m(t) }{N} \right) \left( \frac{ \sigma _{l} (1- \delta _{li}) - m(t) }{N} \right) w'' \left( - 4 \beta J m(t) \sigma _i  \right) G_{N} ( \left\{ F_i \sigma \right\} ; t | t_0 ) \nonumber \\
& & + w \left( - 4 \beta J m(t) \sigma _j  \right) G_{2} ( \sigma _i , - \sigma _j ; t | t_0 ) \nonumber \\ 
& &- \mathrm{Tr} _{ \left\{ \sigma _n \right\} _{n \neq i,j} } 4 \beta J \sigma _j \sum _k \left( \frac{ \sigma _k (1- \delta _{kj}) - m(t) }{N} \right) w' \left( - 4 \beta J m(t) \sigma _j  \right) G_{N} ( \left\{ F_j \sigma \right\} ; t | t_0 ) \nonumber \\
& & + \mathrm{Tr} _{ \left\{ \sigma _n \right\} _{n \neq i,j} } 8 \left( \beta J \right) ^2 \sum _{k,l} \nonumber \\
& & \cdot \left( \frac{ \sigma _k (1- \delta _{kj}) - m(t) }{N} \right) \left( \frac{ \sigma _{l} (1- \delta _{lj}) - m(t) }{N} \right) w'' \left( - 4 \beta J m(t) \sigma _j  \right) G_{N} ( \left\{ F_j \sigma \right\} ; t | t_0 ) \nonumber \\
& & - w \left(  4 \beta J m \sigma _i  \right) G_{2} ( \sigma _i , \sigma _j ; t | t_0 ) \nonumber \\ 
& & - \mathrm{Tr} _{ \left\{ \sigma _n \right\} _{n \neq i,j} } 4 \beta J \sigma _i \sum _k \left( \frac{ \sigma _k (1- \delta _{ki}) - m(t) }{N} \right) w' \left( 4 \beta J m(t) \sigma _i  \right) G_{N} ( \left\{ \sigma \right\} ; t | t_0 ) \nonumber \\ 
& & - \mathrm{Tr} _{ \left\{ \sigma _n \right\} _{n \neq i,j} } 8 \left( \beta J \right) ^2 \sum _{k,l}  \nonumber \\
& & \cdot\left( \frac{ \sigma _k (1- \delta _{ki}) - m(t) }{N} \right) \left( \frac{ \sigma _{l} (1- \delta _{li}) - m(t) }{N} \right) w'' \left( 4 \beta J m(t) \sigma _i  \right) G_{N} ( \left\{ \sigma \right\} ; t | t_0 ) \nonumber \\
& & -w \left(  4 \beta J m(t) \sigma _j  \right) G_{2} ( \sigma _i , \sigma _j ; t | t_0 ) \nonumber \\ 
& & - \mathrm{Tr} _{ \left\{ \sigma _n \right\} _{n \neq i,j} } 4 \beta J \sigma _j \sum _k \left( \frac{ \sigma _k (1- \delta _{kj}) - m(t) }{N} \right) w' \left( 4 \beta J m(t) \sigma _j  \right) G_{N} ( \left\{ \sigma \right\} ; t | t_0 ) \nonumber \\
& & - \mathrm{Tr} _{ \left\{ \sigma _n \right\} _{n \neq i,j} } 8 \left( \beta J \right) ^2 \sum _{k,l} \nonumber \\
& & \cdot \left. \left( \frac{ \sigma _k (1- \delta _{kj}) - m(t) }{N} \right) \left( \frac{ \sigma _{l} (1- \delta _{lj}) - m(t) }{N} \right) w'' \left( 4 \beta J m(t) \sigma _j  \right) G_{N} ( \left\{ \sigma \right\} ; t | t_0 ) \right] .
 \label{MF_dynam2_corr_b}
\end{eqnarray}
In this section, we describe the dependence of $m(t)$ on $t$ explicitly. This equation has the same form as (\ref{MF_dynam2_b}); therefore, the derivation of the time development of $G_2$ is similar to that of $P_2$. 
Note that to calculate the correlation of individual spins $ \left< \delta \sigma _i (t_0) \delta \sigma _j (t) \right>$, we should carefully deal with the difference between focused spins $\sigma _i, \sigma _j$ and others. Consequently, the calculation becomes more complicated than that of $G_2$; therefore, we do not investigate this property in this study.

First, we should calculate the initial condition. From the definition of $\tilde{P} _N$, 
\begin{eqnarray}
\tilde{P} _{N} ( \left\{ \sigma \right\} , t_0 | \left\{ \sigma ' \right\} ; t_0 ) = \delta _{ \left\{ \sigma ' \right\} , \left\{ \sigma \right\} } .
 \label{PN_init}
\end{eqnarray}
Using the above relation, we obtain
\begin{eqnarray}
G_{N} \left( \left\{ \sigma \right\} ; t_0 | t_0 \right) & = & \sum _{i } \frac{1}{N} \sigma _{i} P_N ( \left\{ \sigma \right\} , t_0 ) , \label{GN_init} \\ 
G_{2} \left( \sigma _i , \sigma _j ; t_0 | t_0 \right) & = & \mathrm{Tr} _{ \left\{ \sigma _n \right\} _{n \neq i , j} } \sum _{k } \frac{1}{N} \sigma _{k} P_N ( \left\{ \sigma \right\} , t_0 ) \nonumber \\
 & = & \left( 1 - \frac{2}{N} \right) \sum _{\sigma '} \sigma ' P_3 ( \sigma _i , \sigma _j , \sigma ' , t_0 ) + \frac{1}{N} \left( \sigma _i + \sigma _j \right) P_2 ( \sigma _i , \sigma _j , t_0 ) \nonumber \\
 & = & m(t_0) p_1(\sigma _i , t_0 ) p_1(\sigma _j , t_0 ) + \frac{1}{N} \left\{ \sigma _i + \sigma _j - 2 m(t_0) \right\} p_1(\sigma _i , t_0 ) p_1(\sigma _j , t_0 ) \nonumber \\ & & + \sum _{\sigma '} \sigma ' p_3 ( \sigma _i , \sigma _j , \sigma ' , t_0 ) + o \left( \frac{1}{N} \right) .
 \label{G2_init}
\end{eqnarray}
Substituting (\ref{P2}), (\ref{p2}), (\ref{p3}), and (\ref{alpha3}) into (\ref{G2_init}), the initial condition of $G_2$ is expressed as
\begin{eqnarray}
G_{2} \left( \sigma _i , \sigma _j ; t_0 | t_0 \right) & = & m(t_0) p_1 \left( \sigma _i , t_0 \right) p_1 \left( \sigma _j , t_0 \right) + \frac{\sigma _i + \sigma _j - 2m(t_0)}{N} p_1 \left( \sigma _i , t_0 \right) p_1 \left( \sigma _j , t_0 \right) \nonumber \\ 
& & + \frac{\delta m (t_0) }{4} + \frac{v + 2m (t_0) \delta m (t_0) }{4}  \left( \sigma _i + \sigma _j \right) \nonumber \\
& & + \frac{3m(t_0) \left( v + m(t_0) \delta m (t_0) \right) }{4} \sigma _i \sigma _j  + o \left( \frac{1}{N} \right) . 
\label{G2_init2}
\end{eqnarray}
The $O(1)$ term of $G_2 \left( \sigma _i , \sigma _j ; t_0 | t_0 \right)$ is equal to that of $P_2 \left( \sigma _i , \sigma _j ; t_0 \right) $ multiplied by $m(t_0)$. Hence, considering that the equations describing $G_2$ and $P_2$ have the same forms as mentioned above, the $O(1)$ term of $G_2 \left( \sigma _i , \sigma _j ; t | t_0 \right)$ is expressed as the product of $p_1(\sigma _i , t) p_1(\sigma _j , t) $ and coefficient $m(t_0)$. The same discussions are also applied to $G_3$ and $G_4$; therefore, we can express these functions as
\numparts
\begin{eqnarray}
G_{2} \left( \sigma _i , \sigma _j ; t | t_0 \right) & = & \left( m (t_0 ) + \delta m (t_0 ) \right) p_1(\sigma _i , t) p_1(\sigma _j , t) + g_2 ( \sigma _i , \sigma _j ; t | t_0 ) , \label{G2} \\
G_{3} ( \sigma _i , \sigma _j , \sigma _k ; t | t_0 ) & = & \left( m (t_0 ) + \delta m (t_0 ) \right) p_1(\sigma _i , t) p_1(\sigma _j , t) p_1(\sigma _k , t) + g_3 ( \sigma _i , \sigma _j , \sigma _k ; t | t_0 ) , \label{G3} \\
G_{4} ( \sigma _i , \sigma _j , \sigma _k , \sigma _l ; t | t_0 ) & = & \left( m (t_0 ) + \delta m (t_0 ) \right) p_1(\sigma _i , t) p_1(\sigma _j , t) p_1(\sigma _k , t) p_1(\sigma _l , t) \nonumber \\
& & + g_4 ( \sigma _i , \sigma _j , \sigma _k , \sigma _l ; t | t_0 )  , \label{G4}
\end{eqnarray}
\endnumparts
\begin{eqnarray}
\mathrm{where} \ \ \ g_2 ( \sigma _i , \sigma _j ; t | t_0 ) , g_3 ( \sigma _i , \sigma _j , \sigma _k ; t | t_0 ) , g_4 ( \sigma _i , \sigma _j , \sigma _k , \sigma _l ; t | t_0 ) = O \left( \frac{1}{N} \right)  .
 \label{G2a}
\end{eqnarray}
Here, note that the following relation exists:
\begin{eqnarray}
& & \sum _{\sigma _i , \sigma _j} G_{2} \left( \sigma _i , \sigma _j ; t | t_0 \right) = \sum _{\sigma _i , \sigma _j, \sigma _k} G_{3} \left( \sigma _i , \sigma _j , \sigma _k ; t | t_0 \right) = \sum _{\sigma _i , \sigma _j, \sigma _k , \sigma _l} G_{4} \left( \sigma _i , \sigma _j , \sigma _k , \sigma _l ; t | t_0 \right) \nonumber \\
& = & \mathrm{Tr} _{ \left\{ \sigma \right\} } G_{N} \left( \left\{ \sigma \right\} ; t | t_0 \right) = \mathrm{Tr} _{ \left\{ \sigma ' \right\} } \sum _{i } \frac{1}{N} \sigma ' _{i} P_N ( \left\{ \sigma ' \right\} , t_0 ) \mathrm{Tr} _{ \left\{ \sigma \right\} } \tilde{P} _{N} ( \left\{ \sigma \right\} ; t | \left\{ \sigma ' \right\} , t_0 ) \nonumber \\
& = & \mathrm{Tr} _{ \left\{ \sigma ' \right\} } \sum _{i } \frac{1}{N} \sigma ' _{i} P_N ( \left\{ \sigma ' \right\} , t_0 ) = m (t_0 ) + \delta m (t_0 ) + o \left( \frac{1}{N} \right) .
 \label{G2_Tr}
\end{eqnarray}
In (\ref{G2})-(\ref{G4}), we add $ \delta m (t_0 )$ to the first terms of the right-hand sides; therefore,  
\begin{equation}
\sum _{\sigma _i , \sigma _j} g_{2} \left( \sigma _i , \sigma _j ; t | t_0 \right) = \sum _{\sigma _i , \sigma _j, \sigma _k} g_{3} \left( \sigma _i , \sigma _j , \sigma _k ; t | t_0 \right) = \sum _{\sigma _i , \sigma _j, \sigma _k , \sigma _l} g_{4} \left( \sigma _i , \sigma _j , \sigma _k , \sigma _l ; t | t_0 \right) = o \left( \frac{1}{N} \right) .
 \label{g2_Tr}
\end{equation}
Substituting (\ref{G2}) into (\ref{MF_dynam2_corr_b}), we obtain
\begin{eqnarray}
& & g_2 \left( \sigma _i , \sigma _j , t+ \left. \frac{1}{N} \right| t_0 \right) - g_2 ( \sigma _i , \sigma _j , t | t_0 ) \nonumber \\
& = & + \frac{4 \beta J m (t_0 ) \sigma _i }{N^2} \left[ w' ( 4 \beta J m(t) \sigma _i ) p _1 ( \sigma _i ,t) + w' (-4 \beta J m(t) \sigma _i ) p _1 ( -\sigma _i ,t) \right] \left\{ \frac{m(t)}{2} - \frac{1-2m^2 (t) }{2} \sigma _j \right\} \nonumber \\
& & + \frac{4 \beta J m (t_0 ) \sigma _j }{N^2} \left[ w' ( 4 \beta J m(t) \sigma _j ) p _1 ( \sigma _j ,t) + w' (-4 \beta J m(t) \sigma _j ) p _1 ( -\sigma _j ,t) \right] \left\{ \frac{m(t)}{2} - \frac{1-2m^2 (t) }{2} \sigma _i \right\}\nonumber \\
& & - \frac{m (t_0 ) }{N^2} \bigl\{ p_1 ( - \sigma _i , t ) w \left( - 4 \beta J m(t) \sigma _i  \right) - p_1 ( \sigma _i , t ) w \left(  4 \beta J m(t) \sigma _i  \right) \bigr\} \nonumber \\
& & \cdot \bigl\{ p_1 ( - \sigma _j , t ) w \left( - 4 \beta J m(t) \sigma _j  \right) - p_1 ( \sigma _j , t ) w \left(  4 \beta J m(t) \sigma _j  \right) \bigr\} \nonumber \\
& & + \frac{1}{N} \bigl\{ g_2 ( - \sigma _i , \sigma _j , t | t_0 ) w \left( - 4 \beta J m(t) \sigma _i  \right) + g_2 ( \sigma _i , - \sigma _j , t | t_0 ) w \left( - 4 \beta J m(t) \sigma _j  \right) \bigr. \nonumber \\
& & \bigl. - g_2 ( \sigma _i , \sigma _j , t | t_0 ) w \left(  4 \beta J m(t) \sigma _i  \right) - g_2 ( \sigma _i , \sigma _j , t | t_0 ) w \left(  4 \beta J m(t) \sigma _j  \right) \bigr\} \ \nonumber \\
& & - \frac{4 \beta J }{N} \left[ \sigma _i w' \left( - 4 \beta J m(t) \sigma _i  \right) \sum _{ \sigma ' } \left( \sigma ' - m(t) \right) g_3( -\sigma _i , \sigma _j , \sigma ' ; t | t_0 ) \right. \nonumber \\
& & + \sigma _j w' \left( - 4 \beta J m(t) \sigma _j  \right) \sum _{ \sigma ' } \left( \sigma ' - m(t) \right) g_3( \sigma _i , -\sigma _j , \sigma ' ; t | t_0 ) \nonumber \\
& & + \sigma _i w' \left( 4 \beta J m(t) \sigma _i  \right) \sum _{ \sigma ' } \left( \sigma ' - m(t) \right) g_3( \sigma _i , \sigma _j , \sigma ' ; t | t_0 ) \nonumber \\
& & \left. + \sigma _j w' \left( 4 \beta J m(t) \sigma _j  \right) \sum _{ \sigma ' } \left( \sigma ' - m(t) \right) g_3( \sigma _i , \sigma _j , \sigma ' ; t | t_0 ) \right] \nonumber \\
& & + \left( \frac{ 8 \left( \beta J \right) ^2 m (t_0 ) \left( 1 -m^2 (t) \right) }{N^2} \right) \nonumber \\
& & \cdot \left\{ w'' \left( - 4 \beta J m(t) \sigma _i  \right) p_1( -\sigma _i , t) p_1(\sigma _j , t) + w'' \left( - 4 \beta J m(t) \sigma _j  \right) p_1( \sigma _i , t) p_1( -\sigma _j , t) \right. \nonumber \\
& & \left. - w'' \left( 4 \beta J m(t) \sigma _i  \right) p_1( \sigma _i , t) p_1( \sigma _j , t) - w'' \left( 4 \beta J m(t) \sigma _j  \right) p_1( \sigma _i , t) p_1( \sigma _j , t)  \right\} \nonumber \\
& & + \frac{ 8 \left( \beta J \right) ^2 }{N} \left[ \sum _{ \sigma ' , \sigma '' } \left( \sigma ' - m(t) \right) \left( \sigma '' - m(t) \right) w'' \left( - 4 \beta J m(t) \sigma _i  \right) g_4( -\sigma _i , \sigma _j , \sigma ' , \sigma '' ; t | t_0 ) \right. \nonumber \\
& & + \sum _{ \sigma ' , \sigma '' } \left( \sigma ' - m(t) \right) \left( \sigma '' - m(t) \right) w'' \left( - 4 \beta J m(t) \sigma _j  \right) g_4( \sigma _i , -\sigma _j , \sigma ' , \sigma '' ; t | t_0 ) \nonumber \\
& & - \sum _{ \sigma ' , \sigma '' } \left( \sigma ' - m(t) \right) \left( \sigma '' - m(t) \right) w'' \left(  4 \beta J m(t) \sigma _i  \right) g_4( \sigma _i , \sigma _j , \sigma ' , \sigma '' ; t | t_0 ) \nonumber \\
& & \left. - \sum _{ \sigma ' , \sigma '' } \left( \sigma ' - m(t) \right) \left( \sigma '' - m(t) \right) w'' \left(  4 \beta J m(t) \sigma _j  \right) g_4( \sigma _i , \sigma _j , \sigma ' , \sigma '' ; t | t_0 ) \right] + o \left( \frac{1}{N^2} \right) ,
\label{MF_dynam2body_corr_a}
\end{eqnarray}
This equation has the same form as (\ref{MF_dynam2body_a}) except that the coefficient, $m(t _0) $, is multiplied to the terms not containing $g_2$, $g_3$, nor $g_4$. We introduce $O \left( 1/N \right)$ parameters $\gamma _1$, $\gamma _2$, $\gamma _3$, and $\gamma _4$ and express $g_2$, $g_3$, and $g_4$ by relations similar to (\ref{p2})--(\ref{p4}), respectively.
\numparts
\begin{eqnarray}
g_2 ( \sigma _i , \sigma _j ; t | t_0 ) & = & \frac{\gamma _1 }{4} \left( \sigma _i + \sigma _j \right) + \frac{\gamma _2 + 2m(t) \gamma _1 }{4} \sigma _i \sigma _j , \label{g2} \\
g_3 ( \sigma _i , \sigma _j , \sigma _k ; t | t_0 ) & = & \frac{\gamma _1}{8} \left( \sigma _i + \sigma _j + \sigma _k \right)  + \frac{\gamma _2 + 2m(t) \gamma _1 }{8} \left( \sigma _i \sigma _j + \sigma _j \sigma _k + \sigma _k \sigma _i \right) + \gamma _3 \sigma _i \sigma _j \sigma _k , \label{g3} \\
g_4 ( \sigma _i , \sigma _j , \sigma _k , \sigma _l ; t | t_0 ) & = & \frac{\gamma _1}{16} \left( \sigma _i + \sigma _j + \sigma _k + \sigma _l \right) \nonumber \\
& & + \frac{\gamma _2 + 2m(t) \gamma _1 }{16} \left( \sigma _i \sigma _j + \sigma _i \sigma _k + \sigma _i \sigma _l + \sigma _j \sigma _k + \sigma _j \sigma _l + \sigma _k \sigma _l \right) \nonumber \\
& & + \frac{\gamma _3}{2} \left( \sigma _i \sigma _j \sigma _k + \sigma _i \sigma _j \sigma _l + \sigma _i \sigma _k \sigma _l + \sigma _j \sigma _k \sigma _l \right) + \gamma _4 \sigma _i \sigma _j \sigma _k \sigma _l . \label{g4}
\end{eqnarray}
\endnumparts

Assuming the following relation similar to (\ref{mv2b}):
\begin{eqnarray}
& & \frac{(N-1)(N-2) }{N^2} \left< \sigma _i \sigma _j \sigma _k \right> _G + \frac{3N-2 }{N^2} \left< \sigma _i \right> _G - 3 \frac{ (N-1) \left< \sigma _i \sigma _j \right> _G + 1 }{N} \left< \sigma _i \right> _G + 2 \left< \sigma _i \right> _G ^3 \nonumber \\
& & = O \left( \frac{1}{N^2} \right) , 
 \label{gamma2b}
\end{eqnarray}
where $\left< \bullet \right> _G$ is the weighted average defined as 
\begin{equation}
\left< f( \left\{ \sigma \right\} ; t) \right> _G \equiv \frac{ \mathrm{Tr} _{ \left\{ \sigma \right\} } G_{N} \left( \left\{ \sigma \right\} ; t | t_0 \right) f( \left\{ \sigma \right\} ; t) }{ \mathrm{Tr} _{\left\{ \sigma \right\} } G_{N} \left( \left\{ \sigma \right\} ; t | t_0 \right)  } = \frac{ \left< M (t_0) f( \left\{ \sigma \right\} ; t) \right> }{ \left< M (t_0) \right> }  ,
 \label{Gave}
\end{equation}
we can derive an equation that has the same form as (\ref{alpha3}), expressed as follows:
\begin{equation}
\gamma _3 = \frac{3m(t) \left( \gamma _2 + m(t) \gamma _1 \right) }{8} . 
 \label{gamma3}
\end{equation}
Using (\ref{MF_dynam2body_corr_a}), (\ref{g4}), and (\ref{gamma3}), the time development of $\gamma _1 $ and $\gamma _2$ are expressed as follows:
\begin{eqnarray}
& & \frac{ d \gamma _1}{dt} \nonumber \\
& = & + \frac{8 \beta J m (t_0 ) }{N} \left[ w' ( 4 \beta J m(t) ) p _1 ( 1,t ) + w' (-4 \beta J m(t) ) p _1 ( -1,t ) \right] m(t) \nonumber \\
& & - \gamma _1 \bigl\{ w \left( - 4 \beta J m(t) \right) + w \left( 4 \beta J m(t) \right) \bigr\} \nonumber \\
& & - 4 \beta J  \left[ \gamma _1 \bigl\{ w' \left( - 4 \beta J m(t) \right) + w' \left( 4 \beta J m(t) \right) \bigr\} \right. \nonumber \\
& & \left. + \left( \gamma _2 + m(t) \gamma_1 \right) \bigl\{ - w' \left( - 4 \beta J m(t) \right) + w' \left( 4 \beta J m(t) \right) \bigr\} \right] \nonumber \\
& & - \left( \frac{ 8 \left( \beta J \right) ^2 m (t_0 ) \left( 1 -m^2 (t) \right) }{N} \right) \nonumber \\
& & \cdot \left[ w'' \left( 4 \beta J m(t) \right) - w'' \left( -4 \beta J m(t) \right) + m(t) \left\{ w'' \left( 4 \beta J m(t) \right) + w'' \left( -4 \beta J m(t) \right) \right\} \right] \nonumber \\
& & + 8 \left( \beta J \right) ^2 \left[ -\gamma _2 \left\{ w'' \left( 4 \beta J m(t) \right) - w'' \left( -4 \beta J m(t) \right) \right\} \right. \nonumber \\
& & \left. - m(t) \gamma _2 \left\{ w'' \left( 4 \beta J m(t) \right) + w'' \left( -4 \beta J m(t) \right) \right\} \right] , 
\label{MF_dynam2body_corr_3a2}
\end{eqnarray}
\begin{eqnarray}
& & \frac{d \gamma _2}{dt} \nonumber \\
& = & - \frac{16 \beta J m (t_0 ) }{N} \left[ w' ( 4 \beta J m(t) ) p _1 ( 1,t ) + w' (-4 \beta J m(t) ) p _1 ( -1,t ) \right] \left( 1 - m^2 (t) \right) \nonumber \\
& & - \frac{4 m (t_0 ) }{N} \bigl\{ p_1 ( - 1 , t ) w \left( - 4 \beta J m(t) \right) - p_1 ( 1 , t ) w \left(  4 \beta J m(t) \right) \bigr\} ^2 \nonumber \\
& & - 2 \gamma _2 \bigl\{ w \left( - 4 \beta J m(t) \right) + w \left( 4 \beta J m(t) \right) \bigr\} \nonumber \\
& & - 8 \beta J \left[ \left\{ 4 \gamma ' _3 - m(t) \left( \gamma _2 + m(t) \gamma _1 \right) \right\} \bigl\{ w' \left( 4 \beta J m(t) \right) - w' \left( - 4 \beta J m(t) \right) \bigr\} \right. \nonumber \\
& & \left. + \gamma _2 \bigl\{ w' \left( 4 \beta J m(t) \right) + w' \left( - 4 \beta J m(t) \right) \bigr\} \right] \nonumber \\
& & + 16 \left( \beta J \right) ^2 \left( -4 \gamma ' _4 + m ^2 (t) \gamma _2 \right) \left\{ w'' \left( 4 \beta J m(t) \right) + w'' \left( -4 \beta J m(t) \right) \right\}  ,
\label{MF_dynam2body_corr_3d}
\end{eqnarray}
\begin{eqnarray}
\mathrm{where} \ \ \left\{
\begin{array}{ccc}
\gamma ' _3 & = & 2 \gamma _3  -  \frac{m(t) \left( \gamma _2 + 2m(t) \gamma _1 \right)}{4}  \\
\gamma ' _4 & = & 4 \gamma _4 - 4\gamma _3 m(t) + \frac{m (t) ^2 \left( \gamma _2 + 2m(t) \gamma _1 \right)}{4}  \\
\end{array} 
\right.
\end{eqnarray}
The details of the derivation of these equations are similar to those of (\ref{MF_dynam2body_3a2}) and (\ref{MF_dynam2body_3d}). Substituting (\ref{transition1}) into (\ref{MF_dynam2body_corr_3a2}) and (\ref{MF_dynam2body_corr_3d}), we obtain
\begin{eqnarray}
\frac{ d \gamma _1}{dt} & = & \frac{2 \beta J }{ \cosh ^2 \left( 2 \beta Jm(t) \right) } \cdot \left( \gamma _1 - \frac{m(t_0) m(t) }{N} \right) - \gamma _1 \nonumber \\
& & - \left( \frac{ 4 \left( \beta J \right) ^2 \sinh \left( 2 \beta Jm(t) \right) }{\cosh ^3 \left( 2 \beta Jm(t) \right) } \right) \cdot \left( \gamma _2 + \frac{m(t_0) \left( 1-m^2 (t) \right) }{N} \right) ,
\label{MF_dynam2body_corr_4a2}
\end{eqnarray}
\begin{eqnarray}
\frac{d \gamma _2}{dt} & = & \frac{4 \beta J }{\cosh ^2 \left( 2 \beta Jm(t) \right) } \cdot \left( \gamma _2 + \frac{m(t_0) \left( 1-m^2 (t) \right) }{N} \right) - \frac{m(t_0)}{N} \bigl\{ m(t) - \tanh \left( 2 \beta J m(t) \right) \bigr\} ^2 - 2 \gamma _2 . 
\label{MF_dynam2body_corr_4d}
\end{eqnarray}
To solve the above differential equations, we should derive the initial conditions of $\gamma _1$ and $\gamma _2$. Using (\ref{G2_init2}),  (\ref{G2}), and (\ref{g2}), the initial values of $\gamma _1$ and $\gamma _2$ can be calculated as
\numparts
\begin{eqnarray}
\left. \gamma _1 \right| _{t=t_0} & = \left. \left( m \delta m + v + \frac{1-m^2}{N} \right) \right| _{t=t_0} , \label{G2_init4a} \\ 
\left. \gamma _2 \right| _{t=t_0} & = \left. (m v) \right| _{t=t_0}  .  
\label{G2_init4b}
\end{eqnarray}
\endnumparts
Using (\ref{autoC}), (\ref{G2}), and (\ref{g2}), the autocorrelation of magnetization can be expressed as
\begin{eqnarray}
C(t_0 , t) & \equiv & \frac{1}{N^2} \left\{ \left< M(t_0) M(t) \right> - \left< M(t_0) \right> \left< M(t) \right> \right\} \nonumber \\
& = & \sum _{ \sigma , \sigma ' } \sigma G_{2} \left( t_0 | \sigma  , \sigma ' ; t \right) - \left( m(t_0) + \delta m(t_0 ) \right) \left( m(t) + \delta m(t ) \right) \nonumber \\
& = & \left[ \sum _{ \sigma , \sigma ' } \sigma \left( \left( m(t_0) + \delta m(t_0 ) \right) p_1(\sigma  , t) p_1(\sigma ' , t) + \frac{\gamma _1 }{4} \left( \sigma  + \sigma ' \right) + \frac{\gamma _2 + 2m(t) \gamma _1 }{4} \sigma \sigma '  \right) \right] \nonumber \\ 
& & - \left( m(t_0) + \delta m(t_0 ) \right) \left( m(t) + \delta m(t ) \right) \nonumber \\
& = & \gamma _1 - m(t_0) \delta m(t ) + o \left( \frac{1}{N} \right) ;
\label{autoC2}
\end{eqnarray}
therefore, we can calculate this property directly from the numerical solution of $\gamma _1$.

\begin{figure}[hbp!]
\begin{center}
\begin{minipage}{0.99\hsize}
\includegraphics[width = 8.0cm]{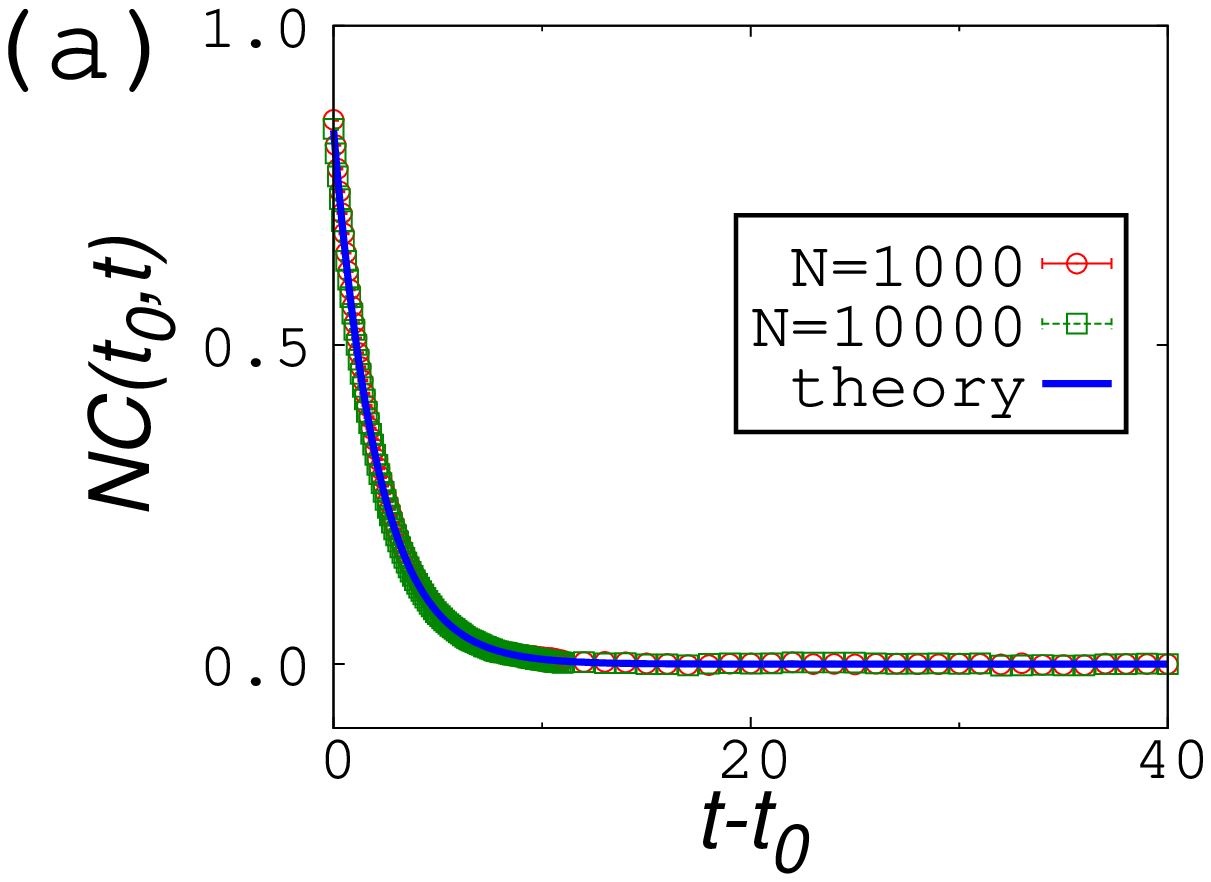}  
\includegraphics[width = 8.0cm]{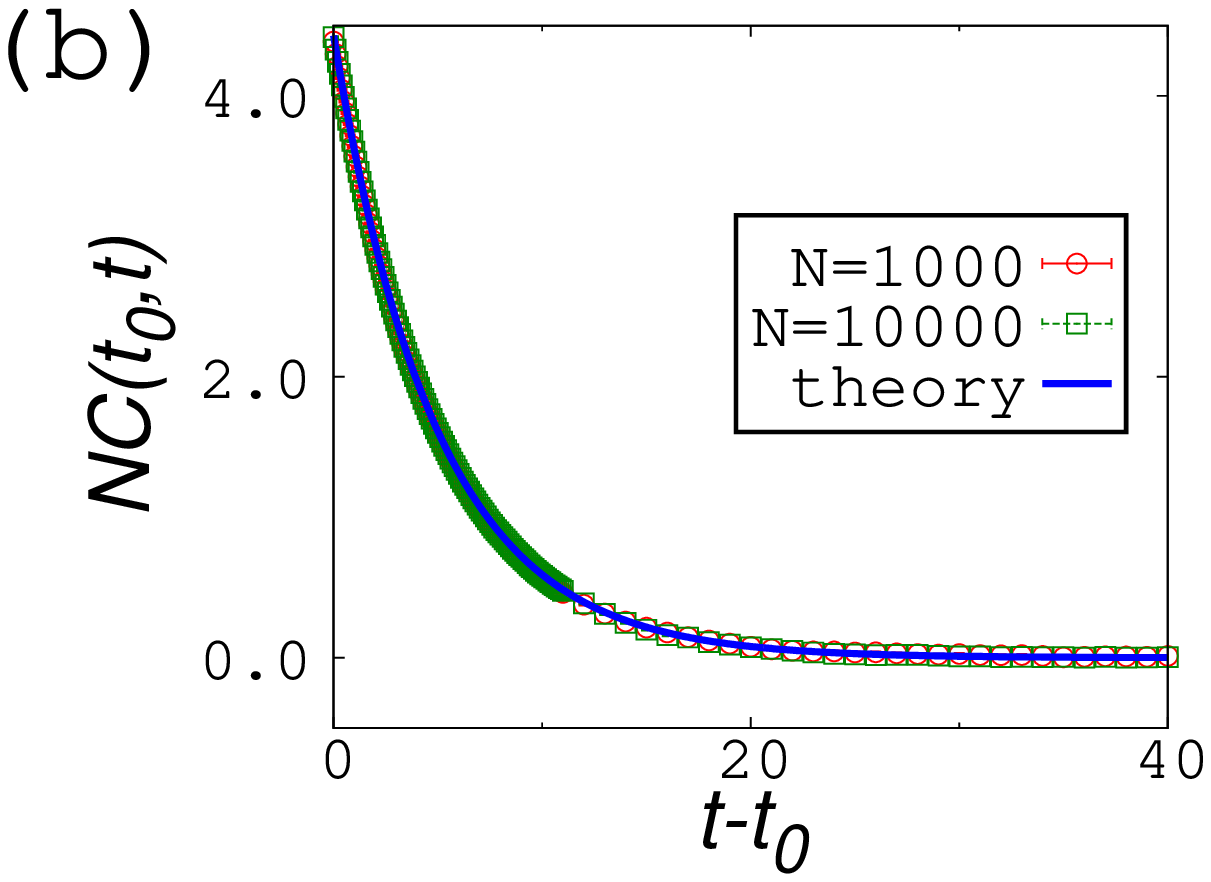} \\ 
\end{minipage}
\end{center}
\caption{(Color online) Autocorrelation of magnetization $NC(t_0 , t) $ at (a)$T = 1.5$ and (b)$T = 2.5$. Red circle and green square points denote results of MCMC simulation at $N = 1000$ and $N = 10000$, respectively, and blue curves are solutions of (\ref{MF_dynam2body_corr_4a2}) and (\ref{MF_dynam2body_corr_4d}). Initial state is perfectly ordered state ($m_0 = 1$), and $t_0$ is set as $t_0 =10$.  }
\label{NCsim}
\end{figure}

Comparison of the solutions of (\ref{MF_dynam2body_corr_4a2}) and (\ref{MF_dynam2body_corr_4d}) and the results of the MCMC simulation is shown in Figure \ref{NCsim}. In this calculation, the initial state is perfectly ordered state $m_0 =1$, and $t_0 $ is set as $t_0 = 10$. Note that $m(t_0)$ in the above discussions is different from $m_0 = m(0)$. These graphs show that the results of the simulations and the differential equations coincide with high accuracy. We also conducted similar calculations for the case of $t_0 = 1$ and 100; however, there are no qualitative differences with Figure \ref{NCsim}. Here, we do not calculate the cases that (\ref{MF_dynam_m2}), (\ref{MF_dynam2body_4a2}), and (\ref{MF_dynam2body_4d}), which are premises of the discussion of this section, become incorrect because of critical phenomena or magnetization reversal.

\section{Summary \label{Summary}}

In this study, we consider the finite-size effect of an infinite-range Ising model by deriving ordinary differential equations describing the $O \left( 1/N \right)$ modifications without using the Fokker--Planck equation. Numerical solutions of these differential equations fit the results of the simulation, unless the fluctuation of magnetization grows large because of critical phenomena or magnetization reversal occurrence. Note that even in these exceptional cases, our equation can describe the time development of the system before the fluctuation grows. As we discuss in \ref{App1}, the derived ordinary differential equations expressed in \ref{2body} provide the same results as the Fokker--Planck equation.

To derive the Fokker--Planck equation for describing the probability distribution of the order parameter, we should understand the number of microscopic states under a given value of the order parameter \cite{PHB89,ML91}. When each variable is more complicated than the Ising spin, this process is difficult, even in the case of infinite-range models. In contrast, our method considers the probability distribution of several spins and calculates its time development without this process. Hence, our method is expected to be applicable to more complicated cases such as the Potts model\cite{ML91,CDLLPS2012,OM13} and the clock model. Such applications should be investigated in future studies.

\appendix

\section{Obtaining time development using Fokker--Planck equation \label{App1} }

In this appendix, we discuss the time development of the system using the Fokker--Planck equation. Under the Glauber dynamics, this equation is expressed as
\begin{eqnarray}
 \frac{\delta P _{\mathrm{whole}} }{\delta t} & = & \frac{\partial }{\partial m' } \left[ \left( m' - \tanh (2 \beta Jm' ) + \frac{2 \beta Jm' }{N \cosh ^2 (2 \beta Jm' )} \right) P _{\mathrm{whole}} \right] \nonumber \\
& & + \frac{1}{N} \frac{\partial ^2}{\partial m ^{\prime 2} } \left[ \left( 1 - m' \tanh (2 \beta Jm' )  \right) P _{\mathrm{whole}} \right] . \label{FP1}
\end{eqnarray}
Here, we let $m' \equiv \left( \sum _i \sigma _i \right) /N$, to distinguish this value from $m = \lim _{N \rightarrow \infty} \left< m' \right>$. In the left-hand side of (\ref{FP1}), the operator, $\frac{\delta}{\delta t}$, is defined by the following relation:
\begin{eqnarray}
 \frac{\delta }{\delta t} f(t) & \equiv & \frac{f \left( t+\frac{1}{N} \right) - f(t)}{\left( \frac{1}{N} \right) } = N \left[ f \left( t+\frac{1}{N} \right) - f(t) \right] , \label{difference_P}
\end{eqnarray}
which converges to the time derivative in the thermodynamic limit, $N \rightarrow \infty$. This quantity is the change in the values during one step of updating(= $1/N$ MCS) divided by the time interval, $1/N$. In previous studies, the time derivative, $\frac{\partial P _{\mathrm{whole}} }{\partial t}$, is in the left-hand side of (\ref{FP1}), instead of the difference, $\frac{\delta P _{\mathrm{whole}} }{\delta t}$. We introduce this modification to consider the contribution of an $O \left( 1/N \right)$ perturbation. Note that the correction term appears when we consider the difference of $\left( f(t) \right) ^2$, i.e.,
\begin{eqnarray}
 \frac{\delta }{\delta t} \left( f(t) \right) ^2 & = & N \left[ \left( f\left( t+\frac{1}{N} \right) \right) ^2 - \left( f(t) \right) ^2 \right] \nonumber \\
& = & 2N f(t) \left[  f \left( t+\frac{1}{N} \right)  - f(t) \right] + N \left[  f \left( t+\frac{1}{N} \right)  - f(t) \right] ^2 \nonumber \\
& = & 2f(t) \frac{\delta f}{\delta t} + \frac{1}{N} \left( \frac{\delta f}{\delta t} \right) ^2 . \label{difference_P2}
\end{eqnarray}
To obtain the differential equations for describing the time development of physical properties, we should consider this $O \left( 1/N \right)$ correction term before taking the limit, $N \rightarrow \infty$.

Integration of (\ref{FP1}) multiplied by $m ^{\prime n}$ yields the time development of moment $\left< m ^{\prime n} \right>$ as follows: 
\begin{eqnarray}
\frac{\delta }{\delta t} \left< m ^{\prime n} \right> & = & -n \left< m ^{\prime n-1} \left( m' - \tanh (2 \beta Jm' ) + \frac{2 \beta Jm' }{N \cosh ^2 (2 \beta Jm' )} \right) \right> \nonumber \\
& & + \frac{n(n-1)}{N} \left< m ^{\prime n-2} \left( 1 - m' \tanh (2 \beta Jm' )  \right) \right> . \label{mn_FP}
\end{eqnarray}
When $n =1$ and 2, (\ref{mn_FP}) is expressed as 
\begin{eqnarray}
\frac{\delta }{\delta t} \left< m ^{\prime} \right> & = & - \left< m' - \tanh (2 \beta Jm' ) + \frac{2 \beta Jm' }{N \cosh ^2 (2 \beta Jm' )} \right> \label{m1_FP} \\
\frac{\delta }{\delta t} \left< m ^{\prime 2} \right> & = & -2 \left< m' \left( m' - \tanh (2 \beta Jm' ) + \frac{2 \beta Jm' }{N \cosh ^2 (2 \beta Jm' )} \right) \right> \nonumber \\
& & + \frac{2}{N} \left< 1 - m' \tanh (2 \beta Jm' )  \right> . \label{m2_FP}
\end{eqnarray}
As previous studies pointed out\cite{AFC10}, these equations themselves need information of higher-order moments. However, if we assume that the probability distribution of $m'$ has a Gaussian form with a sharp peak at $\left< m' \right>$ like the discussions of our main text, we can ignore the contributions of higher-order moments. Letting $\xi \equiv m' - \left< m' \right> $, the moment, $\left< \xi ^n \right>$, is zero unless $n = 2$, because of this assumption. Specifically,
\begin{equation}
\left< \xi ^n \right> = \left\{
\begin{array}{cc}
v + \frac{1-m^2}{N} & \mathrm{if} \ \  n = 2 . \\
0 & \mathrm{otherwise} . \\
\end{array} 
\right.  \label{mn_zero}
\end{equation}
Here, we use (\ref{deltaM}) to relate the value of $\left< \xi ^2 \right>$ with variables $m$ and $v$, which are mentioned in the main text. 
Substituting the relation, $m' = \left< m' \right> + \xi$, and using (\ref{mn_zero}), we can evaluate the average of the hyperbolic function as
\begin{eqnarray}
\left< \tanh (2 \beta Jm' ) \right> & = & \tanh (2 \beta J \left< m' \right> ) - \frac{4 \left( \beta J \right) ^2 \sinh (2 \beta J \left< m' \right> )}{\cosh ^3 (2 \beta J \left< m' \right> )} \left< \xi ^2 \right> + \left< O\left( \xi ^3 \right) \right> \nonumber \\
& = & \tanh (2 \beta J \left< m' \right> ) - \frac{4 \left( \beta J \right) ^2 \sinh (2 \beta J \left< m' \right> )}{\cosh ^3 (2 \beta J \left< m' \right> )} \left( v + \frac{1-m^2}{N} \right) , \label{tanh} \\
\left< m' \tanh (2 \beta Jm' ) \right> & = & \left< m' \right> \tanh (2 \beta J \left< m' \right> ) \nonumber \\
& & + \left[ \frac{2 \beta J}{\cosh ^2 (2 \beta J \left< m' \right> ) } - \frac{4 \left( \beta J \right) ^2 \left< m' \right> \sinh (2 \beta J \left< m' \right> )}{\cosh ^3 (2 \beta J \left< m' \right> )} \right] \left< \xi ^2 \right> + \left< O\left( \xi ^3 \right) \right> \nonumber \\
& = & \left< m' \right>  \left< \tanh (2 \beta J m' ) \right> + \frac{2 \beta J}{\cosh ^2 (2 \beta J \left< m' \right> ) } \left( v + \frac{1-m^2}{N} \right) . \label{mtanh} 
\end{eqnarray}
In the last line of (\ref{mtanh}), (\ref{tanh}) is substituted. Using (\ref{tanh}), (\ref{m1_FP}) can be transformed into
\begin{eqnarray}
 \frac{\delta }{\delta t} \left< m ^{\prime} \right> & = & - \left< m' \right> + \tanh (2 \beta J \left< m' \right> ) - \frac{4 \left( \beta J \right) ^2 \sinh (2 \beta J \left< m' \right> )}{\cosh ^3 (2 \beta J \left< m' \right> )} \left( v + \frac{1-m^2}{N} \right) \nonumber \\
& & - \frac{2 \beta J \left< m' \right> }{N \cosh ^2 (2 \beta J \left< m' \right> )} + O \left( \frac{1}{N^2} \right) . \label{m1_FP2}
\end{eqnarray}
Substituting the notation in the main text that $\left< m' \right> = m + \delta m$, (\ref{m1_FP2}) is expressed as
\begin{eqnarray}
 \frac{\delta m}{\delta t} + \frac{\delta \left(\delta m \right)}{\delta t}  & = & - m + \tanh (2 \beta J m ) \nonumber \\
& & + \left( \frac{2 \beta J }{\cosh ^2 (2 \beta J m )} -1\right) \delta m - \frac{4 \left( \beta J \right) ^2 \sinh (2 \beta J m )}{\cosh ^3 (2 \beta Jm )} \left( v + \frac{1-m^2}{N} \right) \nonumber \\
& & - \frac{2 \beta Jm }{N \cosh ^2 (2 \beta Jm )} + O \left( \frac{1}{N^2} \right) . \label{m1_FP3}
\end{eqnarray}
After taking the thermodynamic limit, this equation is equivalent to the pair of (\ref{MF_dynam_m2}) and (\ref{MF_dynam2body_4a2}). Similarly, (\ref{m2_FP}) can be calculated as
\begin{eqnarray}
 \frac{\delta}{\delta t} \left[ \left< m' \right> ^2 + \left( v + \frac{1-m^2}{N} \right) \right] & = & -2 \left[ \left< m' \right> ^2 + \left( v + \frac{1-m^2}{N} \right) \right] \nonumber \\
& & + 2 \left[ \left< m' \right> \left< \tanh (2 \beta J m' ) \right> + \frac{2 \beta J}{\cosh ^2 (2 \beta J \left< m' \right> ) } \left( v + \frac{1-m^2}{N} \right) \right]  \nonumber \\
& & - \frac{4 \beta J\left< m' \right> ^2 }{N \cosh ^2 (2 \beta J\left< m' \right> )} + \frac{2}{N} \left[ 1- \left< m' \right> \tanh (2 \beta J \left< m' \right> ) \right] + O \left( \frac{1}{N^2} \right) \nonumber \\ 
& = & -2 \left[ \left< m' \right> ^2 + \left( v + \frac{1-m^2}{N} \right) \right] \nonumber \\
& & + 2 \left[ \left< m' \right> \left< \tanh (2 \beta J m' ) \right> + \frac{2 \beta J}{\cosh ^2 (2 \beta J m ) } \left( v + \frac{1-m^2}{N} \right) \right]  \nonumber \\
& & - \frac{4 \beta Jm ^2 }{N \cosh ^2 (2 \beta Jm )} + \frac{2}{N} \left[ 1- m \tanh (2 \beta Jm ) \right] + O \left( \frac{1}{N^2} \right) \label{m2_FP2}
\end{eqnarray}
Here, we used the relation,
\begin{equation}
\left< m ^{\prime 2} \right> = \left< m' \right> ^2 + \left< \xi ^2 \right> = \left< m' \right> ^2 + v + \frac{1-m^2}{N} . \label{m2_supplement}
\end{equation}
Substituting (\ref{MF_dynam_m2}) and (\ref{m1_FP2}) into (\ref{m2_FP2}), we obtain
\begin{eqnarray}
\frac{\delta v}{\delta t} & = & -2v + \frac{4 \beta J}{\cosh ^2 (2 \beta J m ) } \left( v + \frac{1-m^2}{N} \right) - \frac{1}{N} \left( \frac{\delta \left< m' \right>}{\delta t} \right) ^2 + O \left( \frac{1}{N^2} \right) \nonumber \\
& = & -2v + \frac{4 \beta J}{\cosh ^2 (2 \beta J m ) } \left( v + \frac{1-m^2}{N} \right) - \frac{1}{N} \left\{ m - \tanh (2 \beta Jm ) \right\} ^2 + O \left( \frac{1}{N^2} \right) \nonumber \\  \label{m2_FP3}
\end{eqnarray}
Note that the correction term discussed in (\ref{difference_P2}) appears in the last term of the right-hand side of this equation. Taking the limit, $N \rightarrow \infty$, (\ref{m2_FP3}) reduces to (\ref{MF_dynam2body_4d}). Hence, the Fokker--Planck equation yields the same conclusion as our method under the assumption that the probability distribution of magnetization has a sharp Gaussian form.

\section*{Acknowledgments}
The present study was supported by the Grant-in-Aid for Early-Career Scientists (No. 21K13857) from the Japan Society for the Promotion of Science (JSPS). 
A part of the numerical calculations were performed on the Numerical Materials Simulator at the National Institute for Materials Science.


\section*{References}
\begin{thebibliography}{99}


\bibitem{PHB89} Paul W, Hermann D W and Binder K 1989
\textit{J. Phys. A: Math. Gen.} \textbf{22} 3325



\bibitem{AFC10} 
Anteneodo C, Fererro E E and Cannas S A 2010
\textit{J. Stat. Mech.} P07026

\bibitem{MMR10} Mori T, Miyashita S and Rikvold P A 2010
\textit{Phys. Rev. E} \textbf{81} 011135

\bibitem{GMM11} Gudyma I, Maksymov A and Miyashita S 2011
\textit{Phys. Rev. E} \textbf{84} 031126

\bibitem{vK76} van Kampen N G 1976
\textit{Adv. Chem. Phys.} \textbf{34} 245

\bibitem{HGTT84} Hanggi P, Grabert H, Talkner P and Thomas H 1984
\textit{Phys. Rev. A} \textbf{29} 371

\bibitem{LT10} Laferza L F and Toral R 2010
\textit{J. Stat. Phys.} \textit{140} 917

\bibitem{PT18} Peralta A F and Toral R 2018
\textit{Chaos} \textit{28} 106303


\bibitem{CDFR14}
Campa A, Dauxois T, Fanelli D and Ruffo S 2014 
\textit{Physics of Long-Range Interacting Systems} (Oxford: Oxford University Press)

\bibitem{SK68}
Suzuki M and Kubo R 1968
\textit{J. Phys. Soc. Japan} \textbf{24} 51 

\bibitem{CA99}
Chakrabarti B K and Acharyya M 1999
\textit{Rev. Mod. Phys.} \textbf{71} 847 



\bibitem{OYCK00}
Oh S K, Yoon C N, Chung J S and Kang H J 2000
\textit{J. Korean. Phys. Soc.} \textbf{37} 503

\bibitem{ML91}
Mendes J F F and Lage E J S 1991
\textit{J. Stat. Phys.} \textbf{64} 653

\bibitem{CDLLPS2012}
Cuff P, Ding J, Louidor O, Lubetzky E, Peres Y and Sly A 2012
\textit{J. Stat. Phys.} \textbf{149} 432

\bibitem{OM13}
Ostilli M and Mukhamedov F 2013
\textit{Europhys. Lett.} \textbf{101} 60008



\end{thebibliography}
\end{document}